\documentclass[sn-aps]{sn-jnl}

\jyear{2023} %

\theoremstyle{thmstyleone}%
\theoremstyle{thmstyletwo}%

\theoremstyle{thmstylethree}%
\usepackage{graphicx}
\usepackage{float} 
\usepackage{subfigure}
\usepackage{booktabs} 
\usepackage{algorithm}  
\usepackage{algpseudocode} 
\def\C#1#2{C_{#1}^{#2}}

\normalbaroutside 

\begin{document}

\title{
    Hierarchical cycle-tree packing model for optimal K-core attack
}


\author[1,2]{ {Jianwen Zhou}} \email{zhoujianwen@itp.ac.cn}

\author*[1,2,3]{ {Hai-Jun Zhou}}\email{zhouhj@itp.ac.cn}

\affil[1]{\orgdiv{CAS Key Laboratory for Theoretical Physics, Institute of Theoretical Physics}, \orgname{Chinese Academy of Sciences}, \orgaddress{\city{Beijing}, \postcode{100190}, \country{China}}}
\affil[2]{\orgdiv{School of Physical Sciences}, \orgname{University of Chinese Academy of Sciences}, \orgaddress{\city{Beijing}, \postcode{100049}, \country{China}}}
\affil[3]{\orgdiv{MinJiang Cooperative Center for Theoretical Physics}, \orgname{MinJiang University}, \orgaddress{\city{Fuzhou}, \postcode{350108}, \country{China}}}

\abstract{
  The $K$-core of a graph is the unique maximum subgraph within which each vertex connects to $K$ or more other vertices. The optimal $K$-core attack problem asks to delete the minimum number of vertices from the $K$-core to induce its complete collapse. A hierarchical cycle-tree packing model is introduced here for this challenging combinatorial optimization problem. We convert the temporally long-range correlated $K$-core pruning dynamics into locally tree-like static patterns and analyze this model through the replica-symmetric cavity method of statistical physics. A set of coarse-grained belief propagation equations are derived to predict single vertex marginal probabilities efficiently. The associated hierarchical cycle-tree guided attack ({\tt hCTGA}) algorithm is able to construct nearly optimal attack solutions for regular random graphs and Erd\"os-R\'enyi random graphs. Our cycle-tree packing model may also be helpful for constructing optimal initial conditions for other irreversible dynamical processes on sparse random graphs.
}




\maketitle

\section{Introduction}

$K$-core is a fundamental graph concept. For a given graph $G$ formed by a set of vertices and a set of edges between these vertices, its $K$-core is unique and it is the maximum subgraph in which each vertex is connected to at least $K$ other vertices of this same subgraph. The cardinality of a $K$-core is simply the number of vertices in this maximum subgraph. If all the vertices in graph $G$ have at least $K$ attached edges, then the whole graph is a $K$-core. The $K$-core of a general graph can be obtained by deleting from the graph, in an iterative manner, all the vertices which have less than $K$ attached edges together with these associated edges. The final surviving subgraph at the end of this pruning process is then the $K$-core.

The $K$-core concept was initially introduced to study magnetization in lattices~\cite{Pollak-1975,Chalupa-1979} and collective behavior in social networks~\cite{Granovetter-1978, Seidman-1983}. The $K$-core captures some of the key properties of many complex networks and has a wide range of applications~\cite{Kong-2019}. In statistical physics, the kinetically constrained models (e.g., the one introduced by Fredrickson and Andersen~\cite{Fredrickson-1984}) as stochastic models for the liquid--glass transition are intimately related to $K$-core percolation~\cite{Sellitto-2005, Rizzo-2019,Rizzo-2020,Perrupato-2022}. In biology, the $K$-cores of protein-protein interaction networks have been exploited along with phylogenetic analysis to predict the functions of certain proteins~\cite{Altaf-Ul-Amine-2003}, and the innermost $K$-cores are found to be prolific, essential, and evolutionary conserved which might constitute a putative evolutionary backbone of the proteome~\cite{Stefan-2005}. The conscious-to-subliminal transition in the brain was recently studied as a $K$-core percolation process~\cite{Lucini-2019}, and $K$-core decomposition of the brain network suggests that $K$-cores might help protect against cognitive decline in aging~\cite{Stanford-2022}. In ecology, $K$-core is utilized to understand the robustness of bipartite mutualistic networks and to help identify the key species to preserve~\cite{Garcia-Algarra-2017}. In social networks, $K$-core analysis shows that the most efficient spreaders are those located within the most densely connected $K$-core~\cite{Kitsak-2010}. As the $K$-core structure plays important roles in numerous domains, it would be of great interest to perform efficient adversarial attacks to evaluate its robustness and resilience. An interesting optimization goal is to bring the $K$-core to a complete collapse by removing the minimum number of vertices from it. We refer to this minimum-sized vertex set $\Gamma$ as an optimal attack set.

The collapse transition of kinetic $K$-core induced by random removal of vertices has been studied extensively in the literature~\cite{Dorogovtsev-2006,Dorogovtsev-2006b,Azimi-Tafreshi-2014,Baxter-2015,Rui-Jie-2022,Zhao-Zhou-Liu-2013,Zhao-2017}. The vulnerability and resilience of equilibrium $K$-core configurations were also investigated by statistical physics methods~\cite{Wang-etal-2020}. However, the optimal $K$-core attack problem is much more challenging to investigate, and theoretical and algorithmic studies are still relatively sparse~\cite{Guggiola-2015,Schmidt-2019,Zhou-2022,Ma-etal-2023}. For the special case of $K=2$, the problem is known as the minimum feedback vertex set (or decycling) problem. It is one of the classic nondeterministic polynomial-complete (NP-complete) optimization problems, aiming at destroying all the loops of the graph in the most economical way~\cite{Karp-1972,Fomin-etal-2008,Sheng-2002}. Heuristic algorithms inspired by statistical physics, such as the {\tt BPD} (belief propagation guided decimation) and the Min-Sum algorithms~\cite{Zhou-2013,Zhou-2016,Braunstein-2016}, have achieved remarkable success on random graph instances. For general $K$, the {\tt CoreHD} (highest degree in core) algorithm~\cite{Lenka-2016} servers as a simple and effective strategy which iteratively removes one of the vertices with the highest degree in the remaining $K$-core. The performance of {\tt CoreHD} was further improved by the {\tt WN} (weak-neighbor) algorithm~\cite{Schmidt-2019}, which iteratively removes one of the vertices which has high degree but whose neighbors have low degrees. Though very simple, the {\tt WN} algorithm appears to outperform the more sophisticated local algorithm {\tt CI-TM} (collective information in threshold models)~\cite{Sen-2017} and some message-passing algorithms~\cite{Altarelli-2013,Altarelli-2013b}.

The $K$-core attack problem can be understood as an optimal initial condition problem for an irreversible dynamics on a graph~\cite{Reichman-2012,Amin-2014,Dreyer-2009,Zhang-2017}. The local rule of the dynamics is very simple: an inactive vertex will remain inactive (irreversibility), while an active vertex will decay to be inactive if it has less than $K$ active nearest neighbors (threshold and nonlinearity). Constructing a minimum set of initially inactive vertices is known to be a NP-complete challenge~\cite{Dreyer-2009}. To study this optimal initial condition problem by statistical physics methods, a straightforward modeling approach is to map the threshold dynamics into a static Potts model with $T+1$ discrete vertex states $t_i \in\{0,1,\cdots,T\}$, where $t_i=0$ means that vertex $i$ is the seed of the dynamics (i.e., it is inactive from the beginning), and $t_i \geq 1$ means that vertex $i$ becomes inactive at the $t_i$-th time step following the local dynamical rule~\cite{Altarelli-2013,Altarelli-2013b,Zhao-2016}. Although this strategy is conceptually simple, it becomes computationally inefficient as the value of maximum time step $T$ becomes large (a large $T$ value is necessary to guarantee good predictive power). A different modeling strategy was introduced in Ref.~\cite{Zhou-2022}, which viewed the $K$-core collapse dynamics from the static view of cycle-tree packing and compressed all the different time steps $t_i \geq 1$ into a single layer of the packing model. The associated message-passing algorithm of this packing model is very efficient, yet it was able to achieve better solutions than those obtained by the {\tt WN} algorithm.
  
The present work extends the single-layer cycle-tree packing model into a hierarchical multi-layer model, and solves this hierarchical model by the replica-symmetric (RS) cavity method of statistical physics~\cite{Mezard-2009,Zhou-book-2015}. The minimum energy density of the model is evaluated through a set of coarse-grained belief propagation equations, and a message-passing algorithm {\tt hCTGA} (hierarchical cycle-tree guided attack) is implemented for constructing close-to-optimal $K$-core attack sets. Both the theoretical predictions and the algorithmic results of the new model are more refined in comparison with those of the original single-layer model. With the maximum layer number $H$ being only slightly greater than unity (typically $H=3$), the {\tt hCTGA} algorithm consistently beats the {\tt WN} algorithm on all the random graph instances tested in this work. Besides its relevance to the optimal $K$-core attack problem, our hierarchical cycle-tree packing problem has also independent value as a new spin glass model system. The model may be adapted to tackle optimal initial-condition problems for other irreversible dynamical processes on sparse random graphs. It can also help to solve hard random combinatorial optimization problems with a global constraint such as the connected dominating set problem~\cite{Habibulla-2023}.

The next Section~\ref{sec:Model} will describe the hierarchical cycle-tree packing model in detail and outline the RS mean field theory. Section~\ref{sec:Results} reports some of the theoretical and algorithmic results obtained on the ensembles of regular random graphs and Erd\"os-R\'enyi random graphs and compares them with the results achieved in prior efforts~\cite{Guggiola-2015,Schmidt-2019,Zhou-2022}. We conclude this work in Section~\ref{sec:conclusion}. The technical details of the coarse-grained belief propagation equations are deferred to the Appendices~\ref{app:Cfactor}--\ref{app:mfRR}.

\begin{figure}[b]
  \centering
  \includegraphics[width=0.65\textwidth]{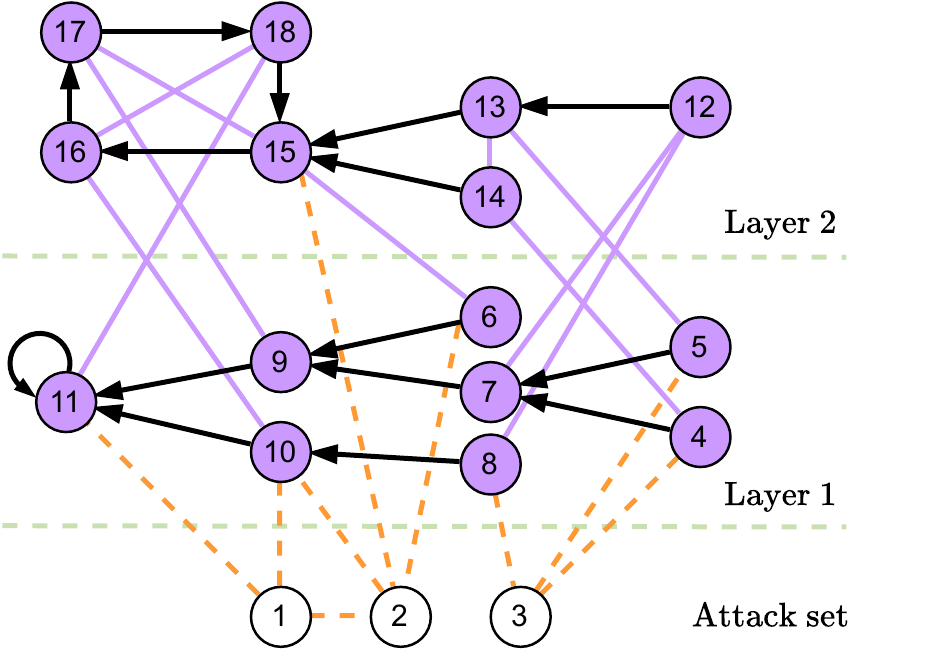}
  \caption{
    Example of hierarchical cycle-tree packing with $H=2$ and $K=3$. The three seed vertices at the lowest layer $h=0$ form the attack set. The edges attached to these seed vertices are drawn as dashed lines. Each normal (non-seed) vertex chooses one of its normal nearest neighbors or itself as the target and this is indicated by an arrow, leading to the formation of a directed tree at layer $h=1$ and a directed cycle-tree at layer $h=2$.
  }
  \label{fig:gexample}
\end{figure}

\section{Model and coarse-grained belief propagation equations}
\label{sec:Model}

We consider a graph $G$ consisting of $N$ vertices and $M$ undirected edges. The vertices are indexed by positive integers $i,j,\ldots \in \{1, 2, \ldots, N\}$, and an edge between two vertices $i$ and $j$ is denoted by $(i,j)$. The neighborhood of vertex $i$ is denoted as $\partial i \equiv \{j:(i,j)\in G\}$ which contains all of its nearest neighbors, and the cardinality of this set is defined as the degree $d_i$ of vertex $i$.

With the optimal $K$-core attack problem in mind,  we now introduce a hierarchical cycle-tree packing model for a general graph $G$. There are $(H+1)$ horizontal layers to accommodate all the $N$ vertices, and every vertex is allocated to one and only one of these layers (Fig.~\ref{fig:gexample}). If vertex $i$ stays at the bottom layer with index $h=0$, we assign it the state $A_i = 0$ and say that it is a seed. If vertex $i$ stays at one of the remaining layers (whose index $h$ is a positive integer, $1 \leq h \leq H$), we say that it is a normal vertex. In this latter case, vertex $i$ needs to choose one of its normal nearest neighbors as its `target'. This target vertex (say $j$) must also be staying at the same layer $h$ as vertex $i$ and furthermore it must \emph{not} choose $i$ as its target. If $j$ is the target of $i$, then vertex $i$ is said to be a `driver' of vertex $j$. We denote this driver-target relation by adding an arrow pointing from $i$ to $j$ on the edge $(i, j)$ and say that `$i$ points to $j$'. Since each normal vertex $i$ is staying at a specific layer with positive index $h$ and it has a unique target vertex $j$, we assign its state as $A_i = j^h$ (the superscript $h$ indicates the specific layer).  A normal vertex $i$ at layer $h$ will choose itself as the target if and only if all its nearest neighbors \emph{at the same layer} $h$ are pointing to $i$ (namely, choosing $i$ as the target). The state of $i$ in this special situation is defined as $A_i = i^h$ and vertex $i$ is then referred to as a root of layer $h$ (there might be no root or be many roots in one single layer).

Consider all the vertices which are staying at the same positive layer $h$. Since each of these vertices is pointing to another vertex of this layer or to itself, these vertices together with the arrow-decorated edges will form one or more tree or cycle-tree components (Fig.~\ref{fig:gexample}). A tree is a connected subgraph containing $n$ vertices and $(n-1)$ edges, while a cycle-tree is a connected subgraph of $n$ vertices and $n$ edges containing a single loop and some tree branches attached to this loop~\cite{Zhou-2013,Zhou-2022}.

Besides the edges decorated by arrows, there are other edges within each single layer of vertices and between two different layers of vertices. In the example of Fig.~\ref{fig:gexample}, these undecorated edges are drawn as solid lines if the vertices at both ends are normal vertices and are drawn as dashed lines if at least one of the end vertices is a seed.

According to the above definitions, each vertex $i$ has a Potts state $A_i$ which can take $\bigl[ (d_i + 1) H + 1\bigr]$ different values. A microscopic configuration is then an $N$-dimensional vector $(A_1, \ldots, A_N)$. It is obvious that not all the possible configurations are valid packing configurations. For any vertex $i$, according to the definition of its state $A_i$, we know that it imposes the following restrictions on the state $A_j$ of its nearest neighbor $j$:
  \begin{enumerate}
  \item[(a)] If $A_i = 0$, then $A_j \neq i^h$ for any $h$.
  \item[(b)] If $A_i = j^h$ with $h \geq 1$, then $A_j = j^h$ or $A_j = k^h$ with $k \in \partial j \backslash i$ (here $\partial j\backslash i$ excludes vertex $i$ from the vertex set $\partial j$).
  \item[(c)] If $A_i = i^h$ with $h \geq 1$, then if $A_j = k^h$ then the vertex index $k$ must be identical to $i$ ($k = i$).
  \end{enumerate}

These three local rules are constraints concerning two neighboring vertices. They ensure that vertices of each positive layer $h$ organize into cycle-tree static patterns. To implement the $K$-core pruning rule in our static model, we also need the following local rule concerning a vertex $i$ and all its nearest neighbors:
\begin{enumerate}
\item[(d)] If vertex $i$ is staying at layer $h \geq 1$ (i.e., $A_i = k^h$ with $k = i$ or $k \in \partial i$),  the summed number ($d_1$) of its nearest neighbors which are staying at layers $h^\prime \geq h$ but are not pointing to $i$ must be less than $K$ ($d_1 <  K$); in addition, if vertex $i$ is staying at layer $h \geq 2$, the summed number ($d_2$) of its nearest neighbors which are staying at layers $h^{\prime\prime} \geq h-1$ but are not pointing to $i$ must be at least $K$ ($d_2 \geq K$).
\end{enumerate}
Such many-body constraints ensure that after all the seed vertices (which stay at the bottom layer) are deleted from graph $G$, the vertices at tree branches of the next layer $h=1$ will all be pruned from the graph following the $K$-core local rule. If the closed loop of a cycle-tree survives this pruning process, we can delete a single vertex from this loop to break it down. After all the vertices in layer $h=1$ are pruned, the vertices in the upper layers $h=2, \ldots, H$ can be pruned consecutively in the same way, leading to the evaporation of the whole $K$-core [Fig.~\ref{fig:gexample}].

Let us define the energy of a microscopic configuration as
\begin{equation}
  E( A_1, \ldots, A_N) = \sum\limits_{i=1}^{N} \delta_{A_i}^0 \; ,
\end{equation}
which is simply the total number of seed vertices. Here $\delta_{x}^{y}$ is the Kronecker symbol such that $\delta_{x}^{y}=1$ if $x$ and $y$ are identical and $\delta_{x}^{y} = 0$ if otherwise. Let us introduce a constraint factor $C_i(A_i, \underline{A}_{\partial i})$ for each vertex $i$ such that $C_{i} = 1$ if the above-mentioned rules (a)--(d) are locally satisfied and $C_{i} = 0$ if at least one of them is violated. Here $\underline{A}_{\partial i} \equiv \{A_j : j \in \partial i\}$ denotes a composite state of the neighboring vertices $j$ of vertex $i$. The explicit expression for the constraint factor $C_i(A_i, \underline{A}_{\partial i})$ could be written down (see Appendix~\ref{app:Cfactor}). Then the partition function of the model is written as
\begin{equation}
  Z(\beta )= \sum _{A_1, \ldots, A_N} \prod\limits_{i=1}^{N}\Bigl[ e^{-\beta \delta_{A_i}^0}
    C_{i}( A_{i} ,\underline{A}_{\partial i}) \Bigr] \; ,
\end{equation}
where $\beta$ is the inverse temperature parameter. Each seed vertex contributes a penalty factor $e^{-\beta}$ to the total Boltzmann weight of a microscopic configuration. The equilibrium probability of observing a specific microscopic configuration is
\begin{equation}
  P( A_1, \ldots, A_N ) = \frac{1}{Z(\beta)}
  \prod\limits_{i=1}^{N}\Bigl[ e^{-\beta \delta _{A_{i}}^{0}}
    C_{i}( A_{i} ,\underline{A}_{\partial i}) \Bigr] \; .
\end{equation}

Because of the $0/1$ constraint factors $C_i$, only those microscopic configurations which satisfy the local constraints of all the $N$ vertices will have nonvanishing contribution to $Z(\beta)$, and this contribution is $e^{-\beta E(A_1, \ldots, A_N)}$. When $\beta$ becomes sufficiently large, the partition function will be contributed almost completely by the minimum-energy cycle-tree packing configurations.

We can solve this hierarchical cycle-tree packing model by the standard RS cavity method of statistical physics. The details of the resulting belief propagation (BP) equations are presented in Appendix~\ref{app:cavmet}, and the coarse-grained BP equations are presented in Appendix~\ref{app:CoBP}. A nice property of the coarse-grained BP equations is that there are only five different coarse-grained states at each positive layer $h \geq 1$, for the vector-formed cavity message $Q_{i\rightarrow j}$ from a vertex $i$ to its nearest neighbor $j$. This property greatly simplifies the numerical computations. Another advantage of this hierarchical cycle-tree packing model is that the maximum layer value $H$ does not need to be large. Our empirical experiences suggest that, in algorithmic applications, a value of $H = 3$ or $H=4$ will be sufficient for many large graph instances (see Section~\ref{sec:hCTGA}).

If $K=2$, our model reduces to the feedback vertex set model of Ref.~\cite{Zhou-2013}. In this special case the exact value of the layer number $H$ becomes irrelevant and we can simply set $H = 1$. The reason is simple: For $K = 2$, after the removal of all the seed vertices, all the edges between the active vertices will be arrow-decorated, except for the few edges between two root vertices at different layers; these connected root vertices must form tree subgraphs due to the local constraints of our model, so we can add an arrow to each such edge pointing from lower layer to higher layer and reduce the total number of hierarchical level to $H = 1$.

When $K \geq 3$, the models with different values of $H$ are different. If we fix $H=1$, our present model reduces to the single-layer cycle-tree packing model of Ref.~\cite{Zhou-2022}. 

\section{Results}
\label{sec:Results}

\subsection{Theoretical predictions for regular random graphs}
\label{sec:RRgraph}

Each vertex in a regular random (RR) graph is connected to the same integer number $D$ of other vertices in the graph. Except for this uniformity of vertex degrees, the connectivity structure of the graph is completely random. Naturally, we assume that the cavity probabilities of the coarse-grained BP equations are independent of the specific edge indices $(i, j)$. Then the coarse-grained BP equations are simplified into $5 H - 1$ coupled equations, which are listed in Appendix~\ref{app:mfRR}. We determine the fixed point of this set of self-consistent equations by the Newton-Raphson iterative method, and then compute the ensemble-averaged mean energy density $\rho$, free energy density $f$, and entropy density $s$ as a function of inverse temperature $\beta$.  

\begin{figure}
  \centering 
  \subfigure[]{
    \includegraphics[width=0.475\linewidth]{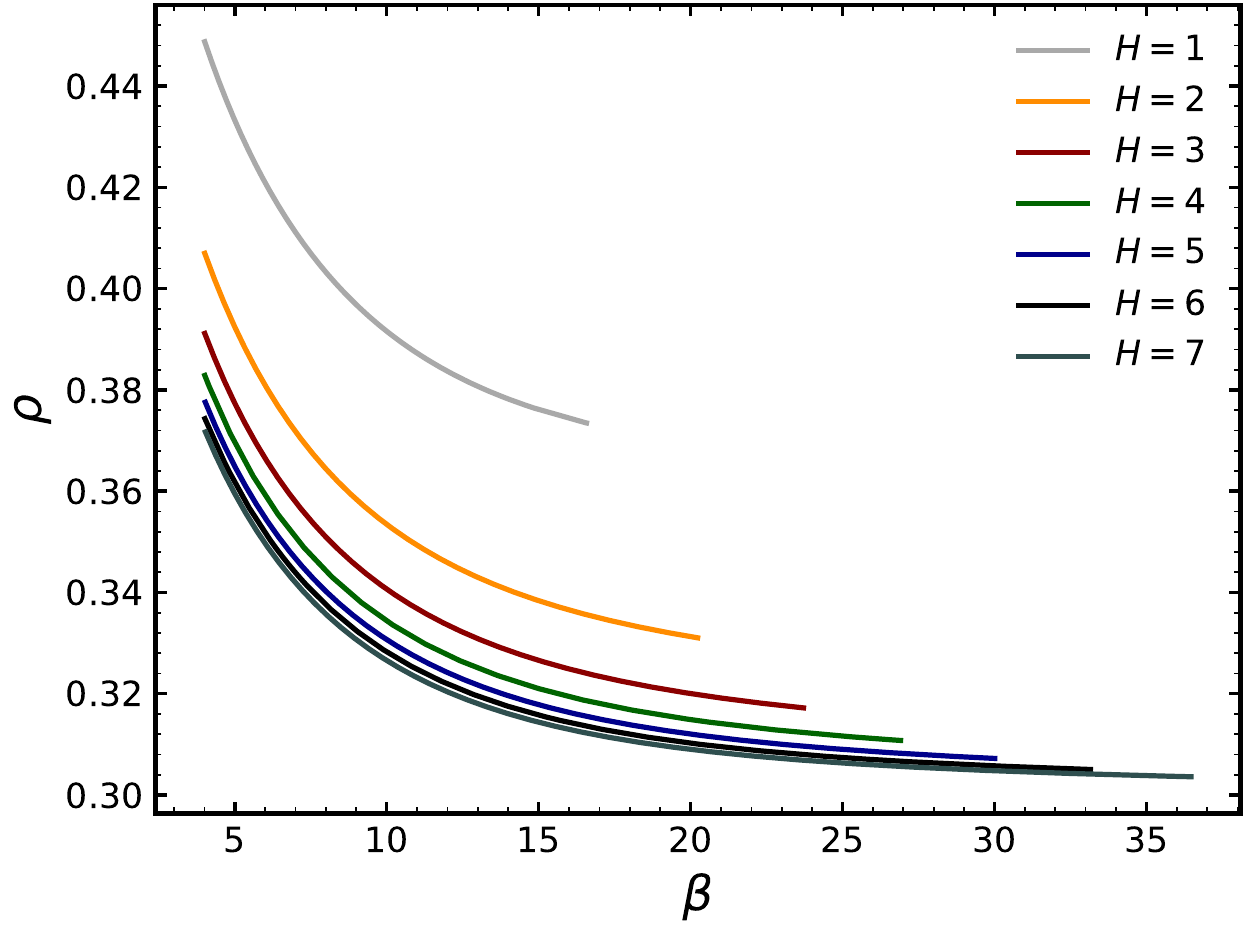}
    \label{fig:RRD7K3:rho}
  }
  \subfigure[]{
    \includegraphics[width=0.475\linewidth]{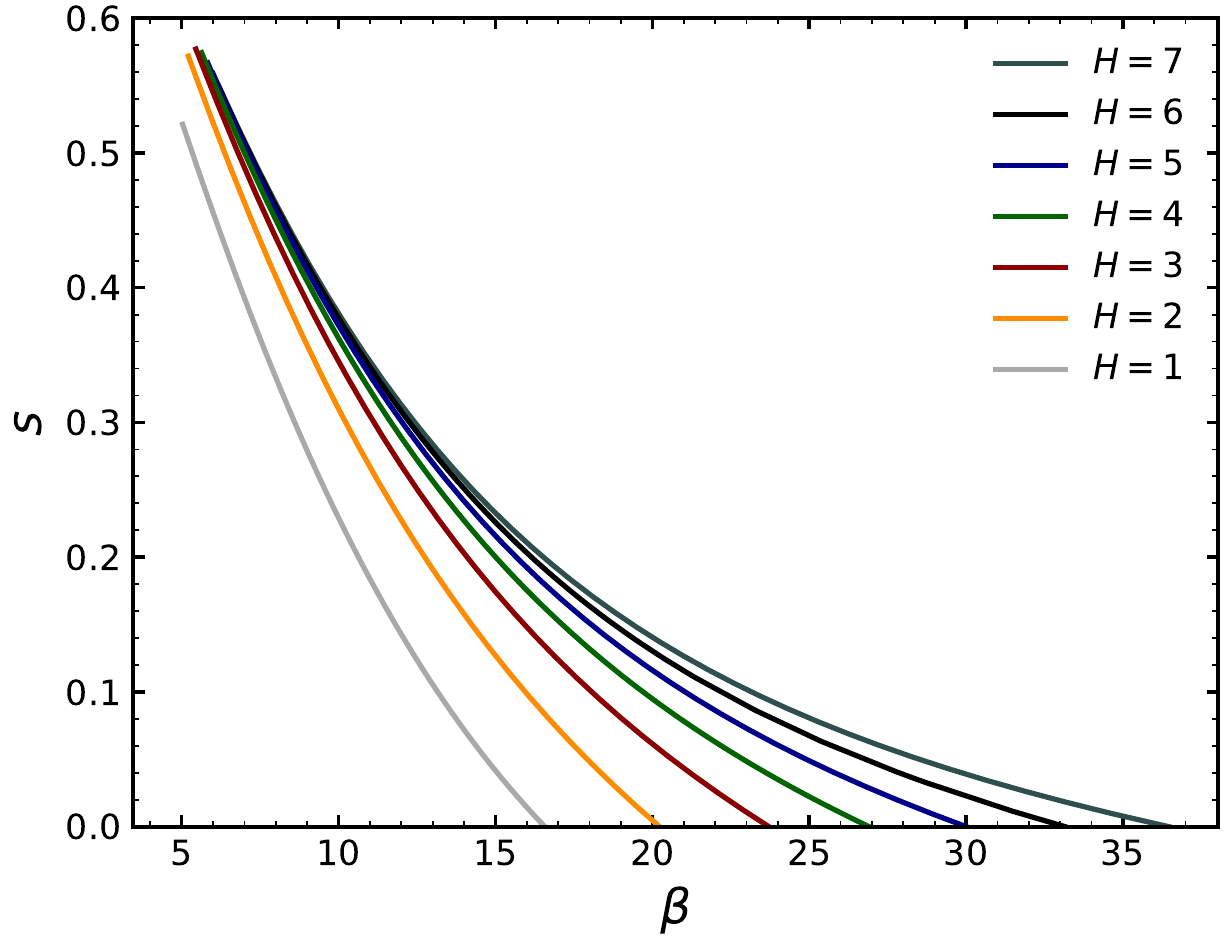}
    \label{fig:RRD7K3:s}
  }
  
  \subfigure[]{
    \includegraphics[width=0.475\linewidth]{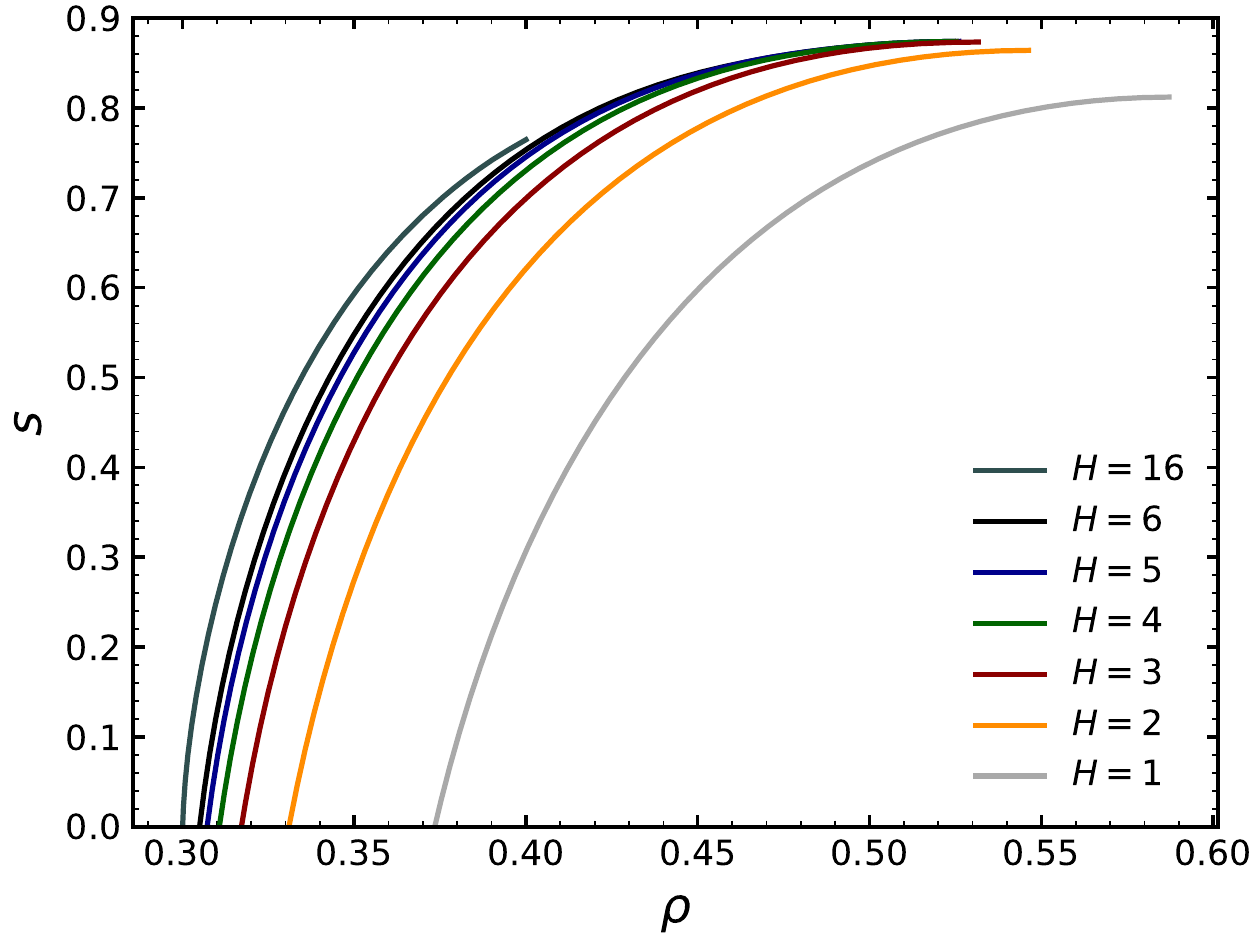}
    \label{fig:RRD7K3:svsrho}
  }
  \subfigure[]{
    \includegraphics[width=0.475\linewidth]{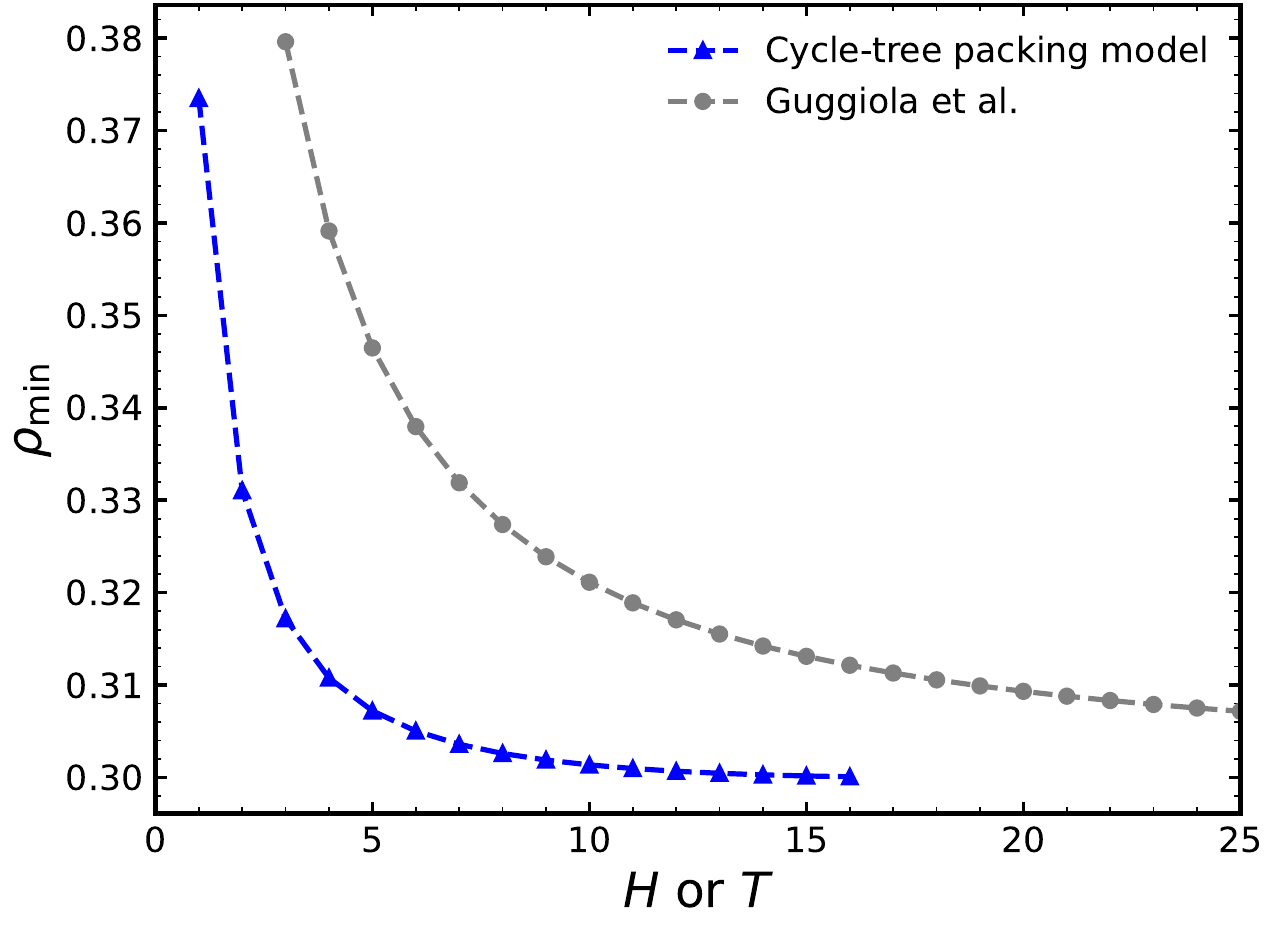}
    \label{fig:RRD7K3:minrho}
  }
  \caption{
    Theoretical predictions for the regular random graph ensemble of degree $D=7$ at different values of layer number $H$. The $K$-core threshold is $K = 3$. (a) Energy density $\rho$ versus inverse temperature $\beta$. (b) Entropy density $s$ versus $\beta$. (c) Entropy density $s$ versus energy density $\rho$. (d) Minimum energy density $\rho_{\textrm{min}}$ versus the value of $H$. As a comparison, the gray points are the results obtained by the model of Ref.~\cite{Guggiola-2015} using the maximum time step $T$ as the hyperparameter.
  }
  \label{fig:RRD7K3}
\end{figure}

Figure~\ref{fig:RRD7K3} summarizes the theoretical results obtained for the RR ensemble of degree $D = 7$ at threshold value $K = 3$. (The results obtained at other values of $D$ and $K \geq 3$ are qualitatively similar.) At each fixed value of level number $H$ the mean energy density $\rho$ [Fig.~\ref{fig:RRD7K3:rho}] and the entropy density $s$ [Fig.~\ref{fig:RRD7K3:s}] both are decreasing functions of inverse temperature $\beta$. The entropy density reaches zero as $\beta$ increases to certain $H$-dependent value and it then becomes negative as $\beta$ further increases. The curves of $s$ versus $\rho$ at different values of $H$ are concave in the region of $s \geq 0$ [Fig.~\ref{fig:RRD7K3:svsrho}]. Negative values of the entropy density indicate that the system enters into the spin glass phase and that the RS mean field theory (which builds on the assumption of only one macroscopic state) is no longer valid at sufficiently low temperatures. Since the true entropy density should be non-negative, within the framework of the RS theory, the value of $\rho$ corresponding to the point of zero entropy density is taken as the predicted minimum energy density and this value is denoted as $\rho_{\textrm{min}}$. As anticipated, this minimum energy density $\rho_{\textrm{min}}$ saturates rather quickly with $H$ [Fig.~\ref{fig:RRD7K3:minrho}]. The limiting value of $\rho_{\textrm{min}}$ at $H \gg 1$ offers a theoretical estimate for the relative size of minimum (optimal) $K$-core attack sets. For $D=7$ and $K=3$ the limiting value of $\rho_{\textrm{min}}$ is $\approx 0.3000$.

For the special case of $K = 2$, the thermodynamic quantities computed at different values of $H \geq 1$ are identical, so the value of $\rho_{\textrm{min}}$ does not depend on $H$. For this special value of $K = 2$, the model of the present work and that of Ref.~\cite{Zhou-2022} reduce to the feedback vertex set model of Ref.~\cite{Zhou-2013}.

From Fig.~\ref{fig:RRD7K3:minrho} we see that the biggest drop in the minimum energy density $\rho_{\textrm{min}}$ occurs as $H$ changes from the minimum value $H = 1$ to $H = 2$. We also observe that the converging speed of $\rho_{\textrm{min}}$ with level number $H$ of the present model is much faster than that of $\rho_{\textrm{min}}$ with the maximum time step $T$ of the conventional propagation model~\cite{Guggiola-2015}. The reason behind this speedy convergence is that our hierarchical packing model has the effect of compressing many time steps of the directional $K$-core pruning (propagation) process into a single hierarchical layer. In our model the coarse-grained cavity message on each edge is a vector of dimension $(5 H - 1)$, while the corresponding cavity message of the propagation model is a vector of dimension $(2 T - 1)$.

The minimum relative size of $K$-core attack sets is predicted to be $0.3000897$ in Ref.~\cite{Guggiola-2015} for $K = 3$ and $D = 7$, by taking the $T \rightarrow \infty$ limit of the propagation model. The predicted value by our present model is slightly lower (the value obtained at $H = 16$ is $0.3000870$). For $K = 2$, the predicted minimum relative sizes are $\rho = 0.25000$ at degree $D=3$, $0.33333$ at $D=4$, $0.37837$ at $D=5$, $0.42199$ at $D=6$ and $0.45892$ at $D=7$, which are either identical to or slightly below the the corresponding theoretical predictions of Ref.~\cite{Guggiola-2015} (see the last column of Table~\ref{tab:RR}). These quantitative comparative results confirm that the hierarchical cycle-tree packing model is a good quantitative model for solving the optimal $K$-core attack problem, even though it is not exactly equivalent to the conventional propagation model of Ref.~\cite{Guggiola-2015}.

The quick convergence of $\rho_{\textrm{min}}$ with $H$ suggests that we will be able to get close-to-optimal solutions with small values of $H$. We now continue to explore this point in the next subsection.

\begin{algorithm}[b]
  \caption{
    Hierarchical Cycle-Tree Guided Attack ({\tt hCTGA}). The main hyperparameter is the maximum number $H$ of layers and the inverse temperature $\beta$. The marginal probability $q_i^0$ of a vertex $i$ belonging to the attack set $\Gamma$ is estimated through Eq.~(\ref{eq:qi0exp}).
  }
  \label{alg:hCTGA}
  \begin{algorithmic}[1]
    \State Read graph $G$ with undirected edges and the vertex set $V$
    \State Initialize coarse-grained BP messages
    \State Initialize the attack set $\Gamma =\emptyset$
    \While{ $V\neq \emptyset$ }
    \State Iterate the coarse-grained BP equations for certain repeats $r$
    \State Calculate the BP marginals $\{q_i^0;i\in V\}$
    \State Choose vertex index $i  = \arg\max_j \bigl( q_j^0 \bigr)$
    \If{ $\textrm{rand()} < q_i^0$} 
    \State Set $\Gamma \leftarrow \Gamma \cup\{i\}$
    \State Set $V \leftarrow V\backslash i$
    \State Set $G \leftarrow K\textrm{-core}(G)$
    \EndIf
    \EndWhile
    \State Return the final attack set $\Gamma$
  \end{algorithmic}
\end{algorithm}

\subsection{Algorithmic results}
\label{sec:hCTGA}

We run the coarse-grained belief propagation equations (listed in Appendix \ref{app:CoBP}) on single graph instances $G$ to estimate the marginal probabilities $q_i^{0}$ of being the seed for all the vertices $i$ [see Eq.~(\ref{eq:qi0exp})]. When the inverse temperature $\beta$ is relatively large, the value of $q_i^{0}$ offers valuable information about the probability of vertex $i$ belonging to an optimal $K$-core attack set. Following the earlier attempt in Ref.~\cite{Zhou-2022}, here we implement a hierarchical cycle-tree guided attack algorithm to construct a close-to-minimum $K$-core attack set $\Gamma$ by sequentially adding vertices to this set under the guidance of the $q_i^0$ marginal probabilities. The pseudo-code is listed in Algorithm~\ref{alg:hCTGA}, where $\textrm{rand(\ )}$ means a random number generator returning a uniformly distributed real value $x \in (0, 1)$, and $K\textrm{-core}(G)$ means returning the $K$-core of a graph $G$. The time complexity of this algorithm is approximately $O(r H N^2)$, where $r$ is the number of repeats of the coarse-grained BP iterations at each decimation step (we simply fix $r = 1$ in our numerical experiments, without waiting for the convergence of the BP iterations). The time complexity could be further reduced to be $O(r H N)$ if we fix a small fraction of the remaining vertices at each decimation step instead of fixing only a single vertex~\cite{Zhou-2022}.

\begin{figure}
  \centering
  \subfigure[]{
    \includegraphics[width=0.475\linewidth]{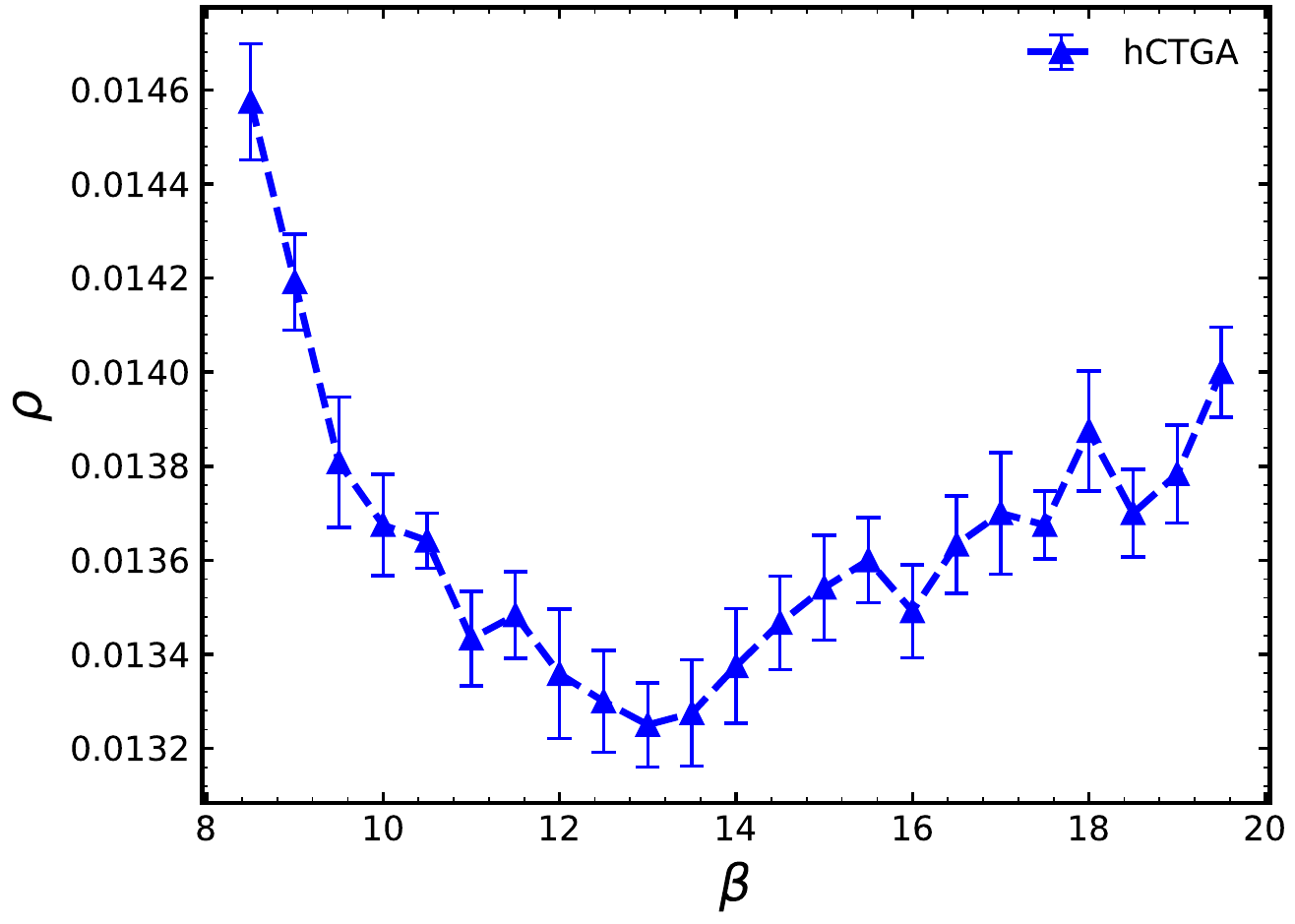}
    \label{fig:hCTGA:beta}
  }
  \subfigure[]{
    \includegraphics[width=0.475\linewidth]{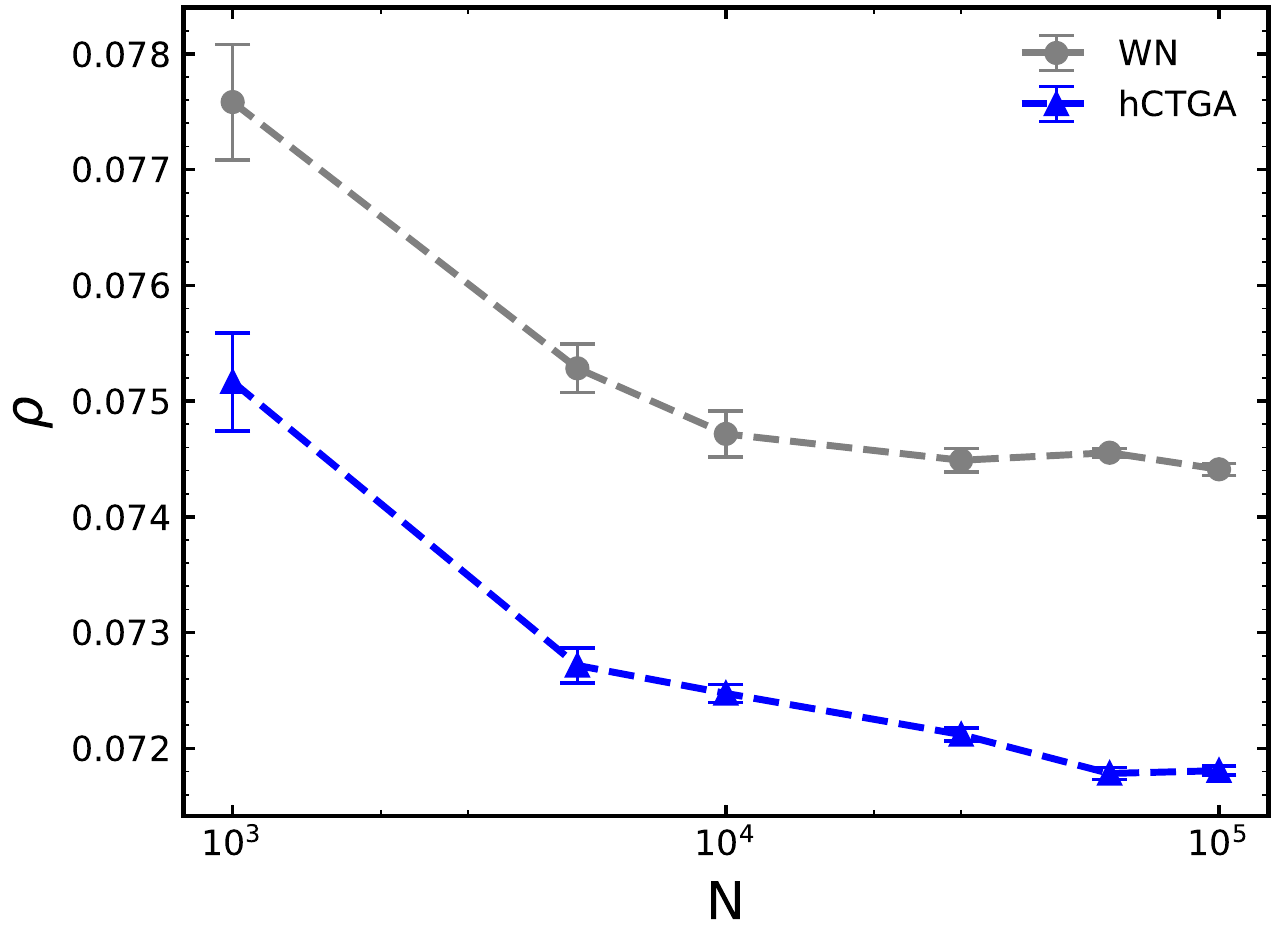}
    \label{fig:hCTGA:N}
  }
  \caption{
    Sensitivity of the results of the {\tt hCTGA} algorithm on the inverse temperature $\beta$ and on the number $N$ of vertices in the graph. The mean values averaged over $12$ regular random graph instances of degree $D$ and size $N$ are shown, and the error bars indicate standard error of the mean. The maximum number $H$ of layers is fixed to $H = 3$.  (a) Relative size $\rho$ of constructed attack set versus $\beta$ on graphs of size $N = 10,000$ and degree $D=6$ with threshold value $K = 5$. (b) Relative size $\rho$ versus graph size $N$, obtained on graphs of degree $D = 4$ with threshold value $K = 3$ and $\beta = 23$. The algorithmic results obtained by the {\tt WN} algorithm~\cite{Schmidt-2019} are also shown as a comparison.
  }
  \label{fig:hCTGA}
\end{figure}

The performance of {\tt hCTGA} has a slight dependence on the inverse temperature and it works best at intermediate $\beta$ values [Fig.~\ref{fig:hCTGA:beta}]. The non-monotonic `V' shape of $\rho$ versus $\beta$ may be due to two competing factors. On the one hand, increasing $\beta$ is beneficial for sampling low-energy configurations; but on the other hand, when $\beta$ becomes large, the system enters the spin glass phase with broken ergodicity and strong frustration. Since the optimal value of $\beta$ might be different for different graph instances,  for a given graph $G$, we run the {\tt hCTGA} algorithm over a range of inverse temperature values and report the minimum-sized attack set $\Gamma$ out of these independent runs.

We also observe that the performance of {\tt hCTGA} improves as the graph size $N$ increases [Fig.~\ref{fig:hCTGA:N}]. This trend may be a manifest of the fact that the mean-field theory behind this algorithm becomes more precise when the typical length of the short loops in the graph increases.

\begin{table}
  \centering
  \begin{tabular}{ccccc}
    \toprule 
    $D$ & $K$  & {\tt hCTGA} & {\tt WN} & prediction \\
    \midrule 
    3 & 2 & 0.2501 & 0.2503  & 0.25000 \\
    4 & 2 & 0.3335 & 0.3378  & 0.33333 \\
      & 3 & 0.0718 & 0.0747  & 0.04628 \\
    5 & 2 & 0.3793 & 0.3964  & 0.37847 \\
      & 3 & 0.1877 & 0.1878  & 0.16668 \\
      & 4 & 0.0264 & 0.0281  & 0.01311 \\
    6 & 2 & 0.4252 & 0.4444  & 0.42262 \\
      & 3 & 0.2621 & 0.2649  & 0.25000 \\
      & 4 & 0.1075 & 0.1085  & 0.07623 \\
      & 5 & 0.0128 & 0.0138  & 0.00572 \\
    7 & 2 & 0.4641 & 0.4836  & 0.46001 \\
      & 3 & 0.3115 & 0.3209  & 0.30009 \\
      & 4 & 0.1801 & 0.1817  & 0.15005 \\
      & 5 & 0.0691 & 0.0694  & 0.04283 \\
      & 6 & 0.0075 & 0.0080  & 0.00310 \\
    \bottomrule 
    \\
  \end{tabular}
  \caption{
    Algorithmic results on the relative size of $K$-core attack set, obtained by the {\tt hCTGA} and the {\tt WN}~\cite{Schmidt-2019} algorithms on regular random graphs of degree $D$. The hyperparameter $H$ of {\tt hCTGA} is $H = 3$ except for the case $D=7$, $K=4$ where it is $H=4$. The algorithmic results are averaged over $12$ independent runs on a single graph instance containing $N = 10,000$ vertices. The last column list the minimum relative size of $K$-core attack sets as predicted by the replica-symmetric mean field theory of Ref.~\cite{Guggiola-2015}.
  }
  \label{tab:RR}
\end{table}
\begin{table}
 \centering
 \begin{tabular}{cccc}
   \toprule 
   D & K  & {\tt hCTGA} & {\tt WN} \\
   \midrule 
   3 & 2 &  0.1353 &    0.1394  \\
   4 & 2 &  0.2138 &    0.2198  \\
     & 3 &  0.0345 &    0.0357  \\
   5 & 2 &  0.2790 &    0.2874  \\
     & 3 &  0.1016 &    0.1019  \\
   6 & 2 &  0.3343 &    0.3438  \\
     & 3 &  0.1630 &    0.1645  \\
     & 4 &  0.0419 &    0.0433  \\
   7 & 2 &  0.3808 &    0.3907  \\
     & 3 &  0.2180 &    0.2196  \\
     & 4 &  0.0976 &    0.0987  \\
     & 5 &  0.0069 &    0.0083  \\
   \bottomrule 
   \\
 \end{tabular}
 \caption{
   Algorithmic results on the relative size of the $K$-core attack set, obtained by the {\tt hCTGA} and the {\tt WN}~\cite{Schmidt-2019} algorithms on Erd\"os-R\'enyi random graphs of mean degree $D$ and size $N = 10,000$. The hyperparameter $H$ of {\tt hCTGA} is fixed to $H = 3$. Each data is the mean value averaged over $12$ random graph instances.
 }
 \label{tab:ER}
\end{table}

The algorithmic results of {\tt hCTGA} on single regular random graph instances are displayed in Table \ref{tab:RR}. The layer number is typically set as $H = 3$ to reduce computational cost. As a comparison we also list the results obtained by the {\tt WN} algorithm of Ref.~\cite{Schmidt-2019} and the theoretical predictions of Ref.~\cite{Guggiola-2015}. The {\tt hCTGA} algorithm outperforms the {\tt WN} algorithm in all the studied graph instances. The {\tt hCTGA} algorithmic results are close to but yet still noticeably exceeding the predicted minimum values. (The performance of {\tt hCTGA} can be further enhanced by increasing $H$, at the cost of an increase in computational time.)

We also run {\tt hCTGA} on Er\"os-R\'enyi random graphs. Such graphs are characterized by a single real-valued parameter, the mean degree $D$. The $(D/2) N$ edges of the graph are chosen uniformly at random out of the $N (N-1)/2$ possible edges. As the results of Table~\ref{tab:ER} demonstrate, our {\tt hCTGA} algorithm (with $H = 3$) again outperforms {\tt WN} on this these random graph instances.

\section{Conclusion and discussion}
\label{sec:conclusion}

The optimal $K$-core attack problem is a prototypical optimization and control task. It is about assigning an optimal initial condition for an irreversible dynamical process to achieve an objective final state many time steps later. In this paper, we generalized the cycle-tree packing model with only a single layer of normal (non-seed) vertices ($H=1$)~\cite{Zhou-2022} to the hierarchical version containing multiple layers of normal vertices ($H \geq 2$), to better address the optimal $K$-core attack problem. We solved this hierarchical cycle-tree packing model by the replica-symmetric mean field theory of spin glass physics, and implemented a message-passing algorithm {\tt hCTGA} to tackle single graph instances. For regular random graphs, the theoretically predicted minimum energy density of this model decreases with the number $H$ of layers and converges to an asymptotic value as $H$ becomes large. The {\tt hCTGA} algorithm has good performance on regular random graphs and  Erd\"{o}s-–R\'{e}nyi random graphs, outperforming the original {\tt CTGA} algorithm with $H=1$~\cite{Zhou-2022} and the simple and striking {\tt WN} heuristic algorithm~\cite{Schmidt-2019}.

The hierarchical cycle-tree packing model is a new member in the zoo of spin glass models. It offered a new perspective on the $K$-core collapse process. The central idea is to represent irreversible $K$-core pruning dynamics by the static and hierarchical cycle-trees patterns in the graph. This strategy of transforming a long-range time-correlated dynamical process into static structural patterns may also be promising for some other hard optimization problems, such as the generalized $K$-core percolation problem on directed graphs~\cite{Zhao-2017}, the directed feedback vertex set and arc set problems~\cite{Zhou-2016,Xu-2017}, and the $K$-core attack by deleting edges~\cite{ZhouBo-2021}. In our model, the hierarchical level $h_i$ of each vertex $i$ is determined by the optimization goal of achieving a percolation of threshold nonlinear dynamics with the minimum initial driving effort. For biological networked systems such as the brain~\cite{Lucini-2019,Wang-2023}, the observed hierarchical organizations in the network might be the result of natural selection over evolutionary time scales. Our hierarchical cycle-tree packing model may also be useful for detecting the hidden hierarchical organization in complex dynamical systems.

In the present work, we carried out the theoretical analysis only at the replica-symmetric level but not at the replica-symmetry-broken levels. The first-step replica-symmetry-breaking mean field theoretical investigation is surely needed to better understand the low-energy landscape of this hierarchical packing model. We expect that the theoretical results will be more precise if the competition effect among different spin glass macroscopic states is taken into account. There are other ways to improve the {\tt hCTGA} algorithm, such as adjusting the value of the inverse temperature $\beta$ at each step of the decimation process, and improving the convergence of the belief propagation iteration dynamics. It will also be interesting to construct the max-sum algorithm~\cite{Altarelli-2013,Altarelli-2013b}, which considers the limit $\beta\to \infty$, and to adapt the reinforcement technique~\cite{Altarelli-2013,Altarelli-2013b}, which uses the noisy information in updating the max-sum equations to force the system to converge to higher energy values. We leave such algorithmic issues for future explorations in combination with practical applications.

\section{Appendix}
\label{sec:app}

\subsection{
  The constraint factor $C_i$ associated with vertex $i$
}
\label{app:Cfactor}

Each vertex $i$ brings a local constraint to its state $A_i$ and the states $\underline{A}_{\partial i}$ of all its nearest neighbors. Here we describe this constraint in some detail.

(1) If vertex $i$ is seed ($A_i=0$), its neighbors are prohibited from pointing to it. Then the constraint factor $C_i$ is
\begin{equation}
  C_{i} (0, \underline{A}_{\partial i} ) =
  \prod\limits_{j\in \partial i} \Bigl( \delta_{A_{j}}^{0}
  + \sum_{h=1}^{H} \bigl[ \delta_{A_{j}}^{j^{h}} + \sum_{l\in \partial j\backslash i} 
    \delta_{A_{j}}^{l^{h}} \bigr] \Bigr ) \; ,
\end{equation}
where $\partial j\backslash i$ contains all the nearest neighbors of vertex $j$, except for vertex $i$. Notice that $C_i = 0$ if $A_j = i^h$ for any $j \in \partial i$ and any $h \geq 1$, otherwise $C_i = 1$.

(2) If vertex $i$ is a root at layer $h=1$ ($A_i=i^1$), then a nearest neighbor $j$ of this vertex is either a seed ($A_j = 0$) or it is staying at the same layer and is pointing to it ($A_j = i^1$), or it is staying at a higher layer $h \geq 2$. The total number of nearest neighbors staying at the higher layers must be less than $K$. Then we can write down the following expression 
\begin{eqnarray}
  C_{i} ( i^{1}, \underline{A}_{\partial i} ) & = &
  \prod\limits_{j\in \partial i}
  \Bigl[ \delta_{A_{j}}^{0} + \delta_{A_{j}}^{i^{1}} +\sum\limits_{h = 2}^{H}
    \bigl( \delta_{A_{j}}^{j^{h}} +\sum\limits_{l\in \partial j\backslash i}
    \delta_{A_{j}}^{l^{h}}\bigr) \Bigr] \ \times
  \nonumber \\
  & & \quad \Theta\Bigl(  K - 1 - \sum\limits_{k\in \partial i}
  \sum\limits_{h=2}^{H}\bigl( \delta_{A_{k}}^{k^{h}} 
  + \sum\limits_{l\in \partial k\backslash i} \delta_{A_k}^{l^{h}} \bigr) \Bigr) \; ,
\end{eqnarray}
where the Heaviside step function $\Theta(x)=1$ if $x\geq 0$ and otherwise $\Theta(x)=0$.

(3) If vertex $i$ is staying at layer $h=1$ and is pointing to vertex $j$ ($A_i=j^1$), then vertex $j$ must also be staying at the same layer and it must not be pointing to $i$. The total number of  other nearest neighbors $k$ of vertex $i$ which are  staying at higher layers or which are staying at the same layer but are not pointing to $i$ must be most $K-2$. Then
\begin{eqnarray}
  & &  C_{i} (j^{1} ,\underline{A}_{\partial i} )  =
  \bigl( \delta_{A_{j}}^{j^{1}} + \sum\limits_{l \in \partial j \backslash i}
  \delta_{A_j}^{l^1} \bigr) \  \times
  \nonumber \\
  & & \quad \quad \prod\limits_{k\in \partial i\backslash j} \Bigl[ \delta_{A_{k}}^{0}
    + \delta_{A_{k}}^{i^{1}} + \sum\limits_{l\in \partial k\backslash i}
    \delta_{A_{k}}^{l^{1}} +\sum\limits_{h=2}^{H} \bigl( \delta_{A_{k}}^{k^{h}}
    + \sum\limits_{l\in \partial k\backslash i} \delta_{A_{k}}^{l^{h}}\bigr) \Bigr]
  \ \times 
  \nonumber \\
  & & \quad \quad \Theta\Bigl( K - 2 - \sum\limits_{m \in \partial i\backslash j}
  \bigl[ \sum\limits_{n\in \partial m\backslash i} \delta_{A_{m}}^{n^{1}}
    +\sum\limits_{h=2}^{H}\bigl( \delta_{A_{m}}^{m^{h}}
    + \sum\limits_{n\in \partial m\backslash i} \delta_{A_m}^{n^{h}}
    \bigr) \bigr] \Bigr) \; .
\end{eqnarray}

(4) If vertex $i$ is a root at layer $h \geq 2$, then all its nearest neighbors at this same layer must be pointing to $i$. The total number of its nearest neighbors staying at layers $t \geq (h - 1)$ must be at least $K$, while the total number of its nearest neighbors staying at layers $s \geq (h + 1)$ must be less than $K$. Therefore for $2\leq h \leq H$, we have
\begin{eqnarray}
  & & C_{i}( i^{h} , \underline{A}_{\partial i} ) =
  \prod\limits_{j\in \partial i}\Bigl[ \delta_{A_{j}}^{0} +\delta_{A_{j}}^{i^{h}}
    + \sum\limits_{t \neq h} \bigl( \delta_{A_{j}}^{j^{t}}
    + \sum\limits_{l\in \partial j\backslash i} \delta_{A_{j}}^{l^{t}}\bigr) \Bigr]
  \  \times
  \nonumber 
  \\
  & & \quad \quad \Theta\Bigl( \sum\limits_{k\in \partial i} \bigl[ \delta_{A_k}^{i^h}
    + \sum\limits_{t>h} \delta_{A_k}^{k^t}
    + \sum\limits_{l\in \partial k\backslash i} ( \delta_{A_k}^{l^{h-1}} 
    +\sum\limits_{t> h} \delta_{A_{k}}^{l^{t}} )\bigr] - K \Bigr)
  \  \times
  \nonumber \\
  & & \quad \quad \Theta\Bigl(  K - 1 - \sum\limits_{k\in \partial i}
  \sum\limits_{t > h} \bigl[ \delta_{A_k}^{k^t} +
    \sum\limits_{l\in \partial k\backslash i} \delta_{A_k}^{l^{t}} \bigr] \Bigr)
  \; .
\end{eqnarray}

(5) If vertex $i$ is staying at layer $h \geq 2$ and it is pointing to vertex $j$, then vertex $j$ must also be staying at layer $h$ and all the other nearest neighbors $k$ of vertex $i$ must not be a root of layer $h$ ($A_k \neq k^h$) or be taking the states $A_k = i^t$ with $t \neq h$ (because such states are meaningless when vertex $i$ is staying at layer $h$). The total number of nearest neighbors of vertex $i$ staying at layers $t \geq h-1$ must be at least $K$, while the total number of nearest neighbors staying at layers $s > h$ must be at most $K-2$. The corresponding expression for $C_i$ is
\begin{eqnarray}
  & & C_{i} (j^{h}, \underline{A}_{\partial i} )  =  \Bigl( \delta_{A_{j}}^{j^{h}} +
  \sum\limits_{l\in \partial j\backslash i} \delta_{A_{j}}^{l^{h}} \Bigr) \times
  \prod\limits_{k\in \partial i\backslash j} \bigl( 1 - \delta_{A_k}^{k^h}
  - \sum\limits_{t \neq h} \delta_{A_k}^{i^t} \bigr) \   \times
  \nonumber  \\
  & & \quad \quad \Theta\Bigl( \sum\limits_{l\in \partial i\backslash j}\bigl[
      \delta_{A_{l}}^{i^{h}}  +\sum\limits_{t > h} \delta _{A_{l}}^{l^t}
      + \sum\limits_{t \geq h-1} \sum\limits_{m\in \partial l\backslash i}
      \delta_{A_{l}}^{m^{t}} \bigr] + 1 - K \Bigr) \ \times
  \nonumber \\
  & & \quad \quad \Theta\Bigl( K- 2 - \sum\limits_{l\in \partial i\backslash j}
  \bigl[ \sum\limits_{t > h} \delta_{A_{l}}^{l^{t}}
    +\sum\limits_{t \geq h} \sum\limits_{m\in \partial l\backslash i}
    \delta_{A_{l}}^{m^{h}} \bigr] \Bigr) \; .
\end{eqnarray}

\subsection{
  Belief propagation equations
}
\label{app:cavmet}

\begin{figure}[b]
  \centering
  \includegraphics[width=0.6\textwidth]{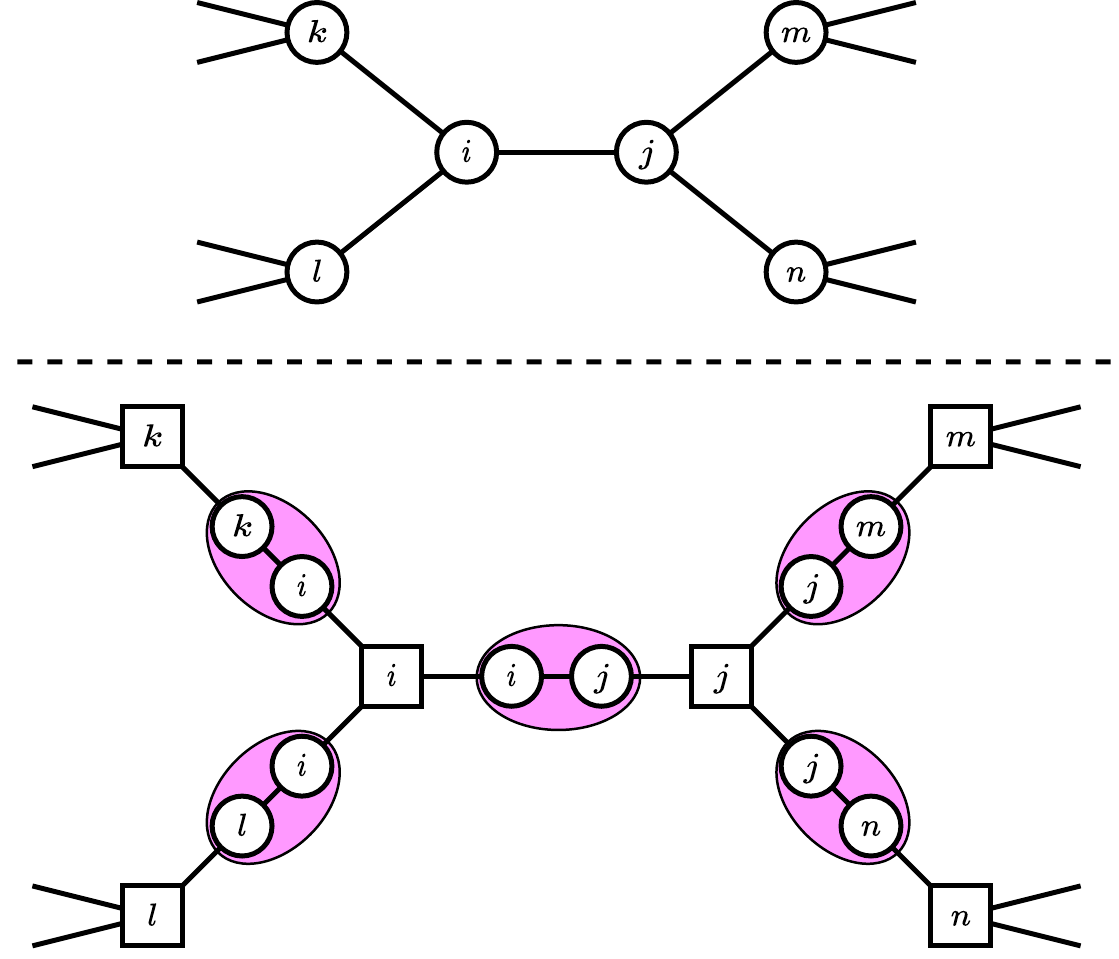}
  \caption{
    Factor-graph representation for deriving the mean field equations. (top) Vertices and edges in the original graph $G$, with $\partial i = \{j, k, l\}$ and $\partial j = \{ i, m, n\}$. (bottom) Variable nodes and factor nodes in the factor graph. Each variable node corresponds to an edge of graph $G$, and each factor node corresponds to the local constraint induced by a vertex of graph $G$. Notice that each variable node is linked to exactly two factor nodes.
  }
  \label{fig:squareandcircle}
\end{figure}

We apply the standard cavity method of statistical physics~\cite{Mezard-2009,Zhou-book-2015} to the hierarchical cycle-tree packing model. By following the same procedure of Refs.~\cite{Zhou-2016,Altarelli-2013}, the self-consistent belief propagation equations can be derived. Because the constraint factors $C_i$ and $C_j$ of two nearest neighboring vertices $i$ and $j$ both involve $A_i$ and $A_j$, it is most convenient to define for each $(i, j)$ a composite state $(A_i, A_j)$. Then the original graph $G$ can be mapped to a factor graph of variable nodes and factor nodes. Each variable node corresponds to an edge $(i, j)$, it has a composite state $(A_i, A_j)$, and it is drawn as an ellipse in the factor graph (Fig.~\ref{fig:squareandcircle}). Each factor node corresponds to the constraint $C_i$ of a vertex $i$, it is drawn as a square and it connects to the variable nodes representing the edges $(i, j)$ with $j\in \partial i$.

We denote by $q_{i\to j}(A_i,A_j)$ the cavity marginal (or belief), the marginal probability of composite state $(A_i,A_j)$ in the absence of the constraint factor $C_j$.  The BP equations on the factor graph are written down as
\begin{equation}
  q_{i\rightarrow j} (A_{i} ,A_{j} )= \frac{1}{z_{i\rightarrow j}}
  \sum\limits_{\underline{A}_{\partial i\backslash j}} e^{-\beta \delta_{A_{i}}^{0}}
  C_{i}(A_{i} ,\underline{A}_{\partial i} ) \prod\limits_{k\in \partial i\backslash j}
  q_{k\rightarrow i} (A_{k} ,A_{i} ) \; , 
  \label{eq:originalBP}
\end{equation}
where $z_{i\to j}$ is a normalization constant to ensure that $\sum_{A_i, A_j} q_{i\rightarrow j}(A_i,A_j) = 1$, and the composite state $\underline{A}_{\partial i \backslash j}\equiv \{A_k: k\in \partial i \backslash j\}$ contains the states $A_k$ of all the nearest neighbors $k$ of vertex $i$, except for that of vertex $j$. The marginal probability of the state $A_i$ of vertex $i$ can be calculated as
\begin{equation}
  q_i^{A_i} =  \frac{1}{z_i} \sum\limits_{\underline{A}_{\partial i}}
  e^{-\beta \delta _{A_{i}}^{0}} C_{i}(A_{i} ,\underline{A}_{\partial i})
  \prod\limits_{k\in \partial i} q_{k\rightarrow i} (A_{k} ,A_{i} ) \; ,
\end{equation}
where $z_i$ is again a normalization constant, and $\underline{A}_{\partial i}\equiv \{A_j: j\in \partial i \}$.

The mean energy density $\rho$, which is simply the mean fraction of seed vertices $i$ (with state $A_i = 0$), is computed through
\begin{equation}
  \rho = \frac{1}{N}\sum\limits_i q_i^0 \; .
\end{equation}
The marginal probability of the composite state $(A_i, A_j)$ of an edge (a variable node of the factor graph, Fig.~\ref{fig:squareandcircle}) is evaluated through
\begin{equation}
  q_{i j}( A_i, A_j) =\frac{1}{z_{i j}} q_{i\rightarrow j}(A_{i} ,A_{j})
  q_{j\rightarrow i} (A_{j} ,A_{i}) \; ,
\end{equation}
with normalization constant $z_{i j}$. The free energy density $f$ of the system  is computed through
\begin{equation}
  f =  -\frac{1}{N \beta}\sum\limits_{i}^{N}\ln z_{i}
  + \frac{1}{N \beta}\sum \limits_{(i,j)\in G} \ln z_{ij} \; .
\end{equation}
The entropy density $s$ at each value of the inverse temperature $\beta$ is simply computed by
\begin{equation}
  s =\beta ( \rho - f ) \; .
\end{equation}
After we obtain the values of $\rho$ and $s$ at different values of $\beta$, we can obtain the relationship of $s$ versus $\rho$, which quantifies the abundance of packing configurations at each fraction of seed vertices. The entropy density needs to be non-negative.

\subsection{The coarse-grained cavity probabilities}

It turns out the BP equations can be simplified considerably by focusing on coarse-grained cavity probabilities. Based on a careful analysis of the BP equations, here we define these coarse-grained probabilities as follows. First for the seed layer $h = 0$, we define
\begin{equation}
  Q_{i\rightarrow j}^{0} \equiv \frac{1}{1+H d_{j}}\Bigl[ q_{i\rightarrow j} (0,0)
    + \sum _{h=1}^{H} \bigl[ q_{i\rightarrow j}( 0,j^{h} ) + 
      \sum\limits_{l\in \partial j\backslash i} q_{i\rightarrow j}( 0, l^{h} ) \bigr]
    \Bigr] \; .
\end{equation}
Here $d_j$ is the degree of vertex $j$. We could interpret $Q_{i\rightarrow j}^0$ as the marginal probability of vertex $i$ being a seed in the absence of the constraint factor $C_j$.

Second, for the lowest active layer $h = 1$, we define the probabilities $Q_{i\rightarrow j}^{x, h=1}$ with inter index $x \in \{1, \ldots, 5\}$ as
\begin{eqnarray}
  Q_{i\rightarrow j}^{1,1} & \equiv & \frac{1}{2} \Bigl[ q_{i\rightarrow j}( i^{1} ,0 )
    + q_{i\rightarrow j}( i^{1}, i^{1}) + \sum\limits_{k\in \partial i\backslash j} \bigl[
      q_{i\rightarrow j} ( k^{1}, 0 ) + q_{i\rightarrow j}( k^{1} ,i^{1} ) \bigr]
    \Bigr] \; ,
  \nonumber \\
  & & 
  \\
  Q_{i\rightarrow j}^{2, 1} & \equiv &  0 \; ,
  \\
  Q_{i\rightarrow j}^{3, 1} & \equiv & \frac{1}{(H - 1) d_{j}}\sum _{t = 2}^{H}\Bigl[
    q_{i\rightarrow j}( i^{1} ,j^{t}) +\sum\limits_{l\in \partial j\backslash i}
    q_{i\rightarrow j}( i^{1} ,l^{t}) \Bigr] \; ,
  \\
  Q_{i\rightarrow j}^{4,1} & \equiv & \frac{1}{d_{j}}
  \Bigl[ q_{i\rightarrow j}(j^{1}, j^{1}) + \sum\limits_{l\in \partial j\backslash i}
    q_{i\rightarrow j}( j^{1} ,l^{1} ) \Bigr] \; ,
  \\
  Q_{i\rightarrow j}^{5,1} & = & \frac{1}{H d_{j} -1}
  \sum\limits_{k\in \partial i\backslash j}\Bigl[ \sum\limits_{t=2}^{H}
    q_{i\rightarrow j}( k^{1} ,j^{t}) + \sum\limits_{t = 1}^{H}
    \sum\limits_{l\in \partial j\backslash i} q_{i\rightarrow j}( k^{1} ,l^{t}) \Bigr] \; .
\end{eqnarray}
We see that $Q_{i\rightarrow j}^{1, 1}$ corresponds to the situation of vertex $i$ staying in layer $h=1$ and vertex $j$ being a seed or being targeted at $i$; $Q_{i\rightarrow j}^{3, 1}$ corresponds to vertex $i$ being a root of layer $h=1$ and vertex $j$ staying at a higher layer; $Q_{i\rightarrow j}^{4, 1}$ corresponds to vertex $i$ staying at layer $h=1$ and pointing to vertex $j$ of the same layer; and $Q_{i\rightarrow j}^{5, 1}$ corresponds to the situation of vertex $i$ pointing to a nearest neighbor $k$ other than $j$ and vertex $j$ staying at a higher layer or being staying at the same layer $h=1$ but are \emph{not} pointing to $i$.

Third, for all the upper active layers with $h = 2, \ldots, H$, we define
\begin{eqnarray}
  Q_{i\rightarrow j}^{1,h} & = & \frac{1}{(h-2) d_{j} +2}
  \Bigl[ q_{i\rightarrow j}( i^{h} ,0 ) + \sum\limits_{k\in \partial i\backslash j}
    q_{i\rightarrow j}( k^{h} , 0 ) \nonumber \\
    &  & \quad \quad 
    + \sum\limits_{t=1}^{h-1} \bigl[ q_{i\rightarrow j}( i^{h}, j^{t})
      + \sum\limits_{k\in \partial i\backslash j} q_{i\rightarrow j}( k^{h}, j^{t}) \bigr]
    \nonumber \\
    & & \quad \quad 
    +\sum\limits_{t \geq 1}^{h-2} \sum\limits_{l\in \partial j\backslash i}
    \bigl[ q_{i\rightarrow j}( i^{h} , l^{t} ) +
      \sum\limits_{k\in \partial i\backslash j} q_{i\rightarrow j}( k^{h} , l^{t} )
      \bigr]  \Bigr] \; ,
  \\
  Q_{i\rightarrow j}^{2,h} & = & \frac{1}{d_{j}} \Bigl[ q_{i\rightarrow j}( i^{h} , i^{h})
    +\sum\limits_{l\in \partial j\backslash i} q_{i\rightarrow j}( i^{h} , l^{h-1})
    \nonumber \\
    & & \quad \quad
    +\sum\limits_{k\in \partial i\backslash j} \bigl[ q_{i\rightarrow j}( k^{h} ,i^{h})
      + \sum\limits_{l\in \partial j\backslash i} q_{i\rightarrow j}( k^{h} ,l^{h-1})
      \bigr] \Bigr] \; ,
  \\
  Q_{i\rightarrow j}^{3,h} & = & \frac{1}{(H - h) d_{j}}\sum\limits_{t= h+1}^{H}
  \Bigl[ q_{i\rightarrow j}( i^{h} , j^{t}) +\sum\limits_{l\in \partial j\backslash i}
    q_{i\rightarrow j}( i^{h} ,l^{t}) \Bigr] \; ,
  \\
  Q_{i\rightarrow j}^{4,h} & = & \frac{1}{d_{j}}\Bigl[ q_{i\rightarrow j}( j^{h} ,j^{h} )
    + \sum\limits_{l\in \partial j\backslash i} q_{i\rightarrow j}( j^{h} ,l^{h} )
    \Bigr] \; ,
  \\
  Q_{i\rightarrow j}^{5,h} & = & \frac{1}{(H - h +1) d_{j} -1}
  \sum\limits_{k\in \partial i\backslash j}\Bigl[ \sum\limits_{t = h+1}^H
    q_{i\rightarrow j}( k^{h} ,j^{t})
    \nonumber \\
    & & \quad \quad
    + \sum\limits_{t = h}^H \sum\limits_{l\in \partial j\backslash i}
    q_{i\rightarrow j}( k^{h} ,l^{t}) \Bigr] \; .
\end{eqnarray}
The summed probability $(Q_{i\rightarrow j}^{1, h} + Q_{i\rightarrow j}^{2, h})$ corresponds to the situation of vertex $i$ staying at layer $h$, and vertex $j$ staying at a lower layer $t$ (with $0 \leq t \leq h-1$) or staying the same layer $h$ and pointing to vertex $k$;  $Q_{i\rightarrow j}^{2, h}$ corresponds to vertex $i$ staying at layer $h$ and vertex $j$ pointing to $i$ or being staying at the layer $t$ immediately below ($t = (h - 1)$);  $Q_{i\rightarrow j}^{3, h}$ corresponds to vertex $i$ being a root of layer $h$ and vertex $j$ staying at a higher layer; $Q_{i\rightarrow j}^{4, h}$ corresponds to vertex $i$ staying at layer $h$ and pointing to vertex $j$ of the same layer; and $Q_{i\rightarrow j}^{5, h}$ corresponds to the situation of vertex $i$ staying at layer $h$ and pointing to a nearest neighbor $k$ other than $j$ and vertex $j$ staying at a higher layer $t$ or staying at the same layer $h$ but are \emph{not} pointing to $i$.

\subsection{
  The coarse-grained belief propagation equations
}
\label{app:CoBP}

Based on the BP equation  (\ref{eq:originalBP}) for the $q_{i\rightarrow j}(A_i, A_j)$ probabilities, we can derive the BP equations for the coarse-grained $Q_{i\rightarrow j}^0$ and $Q_{i\rightarrow j}^{x, h}$ probabilities. We list these coarse-grained BP message-passing equations here. 

First, for the seed layer $h=0$, the self-consistent equation is
\begin{equation}
  Q_{i\rightarrow j}^{0} = \frac{1}{z_{i\rightarrow j}} e^{-\beta }
  \prod\limits_{k\in \partial i\backslash j}\Bigl[ Q_{k\rightarrow i}^{0} +
    Q_{k\rightarrow i}^{1, 1} + \sum\limits_{h = 2}^H Q_{k\rightarrow i}^{1,h} \Bigr] \; .
  \label{eq:Qij0}
\end{equation}

Second, for the next lowest layer $h = 1$, besides the trivial expression $Q_{i\rightarrow j}^{2, 1} = 0$, we have
\begin{subequations}
  \begin{align}
    Q_{i\rightarrow j}^{1,1} & = \frac{1}{z_{i\rightarrow j}} \Bigl[
      G_i^{1}\bigl( K-1,\; \partial i\backslash j \bigr)
      + \sum\limits_{k \in \partial i\backslash j} Q_{k\rightarrow i}^{1, 1}
      G_i^{2}\bigl( K - 2,\;  \partial i\backslash \{j, k\} \bigr) \Bigr] \; ,
    \\
    Q_{i\rightarrow j}^{3,1} & = \frac{1}{z_{i\rightarrow j}}
    G_i^{1}\bigl( K-2,\;  \partial i\backslash j \bigr) \; ,
    \\
    Q_{i\rightarrow j}^{4,1} & = \frac{1}{z_{i\rightarrow j}}
    G_i^{2}\bigl( K-2,\; \partial i\backslash j \bigr) \; ,
    \\
    Q_{i\rightarrow j}^{5,1} & = \frac{1}{z_{i\rightarrow j}}
    \sum\limits_{k\in \partial i\backslash j} Q_{k\rightarrow i}^{1,1}
    G_i^{2}\bigl( K - 3,\; \partial i\backslash \{j, k\} \bigr) \; .
  \end{align}
\end{subequations}
In the above iterative equations, the function $G_i^1(n, V)$, with $n$ being a positive integer and $V$ being a set of vertices, is defined as
\begin{eqnarray}
  G_i^1( n, V ) = \sum\limits_{U \subset V, |U| \leq n}\;  \prod\limits_{j\in U}
  \Bigl( \sum\limits_{h = 2}^{H} Q_{j\rightarrow i}^{1, h} \Bigr)
  \prod\limits_{k\in V \backslash U} \Bigl( Q_{k\rightarrow i}^0 +
  Q_{k\rightarrow i}^{4, 1} \Bigr) \; .
\end{eqnarray}
Here the summation is over all the subsets $U$ of vertex set $V$ under the constraint of the cardinality of $U$ should not exceed $n$; and $V\backslash U$ means the subset of $V$ that is complementary to subset $U$. Similarly the function $G_i^2( n, V)$ is defined as
\begin{eqnarray}
  G_i^2( n, V ) & = & \sum\limits_{U \subset V, |U| \leq n} \; \prod\limits_{j\in U}
  \Bigl( Q_{j\rightarrow i}^{5, 1} + Q_{j\rightarrow i}^{2, 2} +
  \sum\limits_{h =3}^H Q_{j\rightarrow i}^{1, h} \Bigr)
  \nonumber \\
  & & \quad \quad \quad \quad \quad \times 
  \prod\limits_{k\in V \backslash U} \Bigl( Q_{k\rightarrow i}^0 +
    Q_{k\rightarrow i}^{4, 1} \Bigr) \; .    
\end{eqnarray}

Third, for all the upper layers with $h \geq 2$, we have 
\begin{subequations}
  \begin{align}
    Q_{i\rightarrow j}^{1, h} & =  \frac{1}{z_{i \rightarrow j}} \Bigl[
      G_{i}^{3, h}\bigl( K-1, \; K,\; \partial i\backslash j \bigr)
      \nonumber \\
      & \quad \quad \quad \quad \quad
      + \sum\limits_{k\in \partial i\backslash j}  Q_{k\rightarrow i}^{2,h} \;
      G_{i}^{4, h}\bigl(K-2,\; K-1,\; \partial i\backslash \{j, k\} \bigr)
      \Bigr] \; ,
    \\
    Q_{i\rightarrow j}^{2, h} & = \frac{1}{z_{i\rightarrow j}} \Bigl[
      G_{i}^{3, h}\bigl(K - 1,\; K - 1, \; \partial i\backslash j \bigr)
      \nonumber \\
      & \quad \quad \quad \quad \quad
      + \sum\limits_{k\in \partial i\backslash j} Q_{k \rightarrow i}^{2,h}\; 
      G_{i}^{4, h}\bigl( K-2,\; K-2,\; \partial i\backslash \{j, k\} \bigr)
      \Bigr] \; ,
    \\
    Q_{i\rightarrow j}^{3, h} & = \frac{1}{z_{i\rightarrow j}}
    G_{i}^{3, h}\bigl( K - 2,\; K - 1,\; \partial i\backslash j \bigr)
    \; ,
    \\ 
    Q_{i\rightarrow j}^{4, h} & = \frac{1}{z_{i\rightarrow j}}
    G_{i}^{4, h}\bigl( K - 2 , \; K - 1 ,\; \partial i\backslash j \bigr) \; ,
    \\
    Q_{i\rightarrow j}^{5, h} & = \frac{1}{z_{i\rightarrow j}}
    \sum\limits_{k \in \partial i\backslash j} Q_{k\rightarrow i}^{2, h} \;
    G_{i}^{4, h}\bigl( K - 3 ,\; K - 2 ,\; \partial i\backslash \{j, k\} \bigr)
    \; .
  \end{align}
  \label{eq:coBPh2p}
\end{subequations}
Here the functions $G_{i}^{3, h}(m, n, V)$ and $G_{i}^{4, h}(m, n, V)$ are shorthand notations for the following two longer expressions:
\begin{subequations}
  \begin{align}
    G_{i}^{3, h}\bigl(x , y , V) & = \sum\limits_{U_1 \subset V}^{|U_1| \leq x} \;
    \sum\limits_{U_2 \subset V\backslash U_1}^{|U_1\cup U_2| \geq y} \;
    \prod\limits_{k \in U_1}\Bigl[ \sum\limits_{t = h+1}^H Q_{k\rightarrow i}^{1, t}\Bigr]
    \prod\limits_{l\in U_2} \Bigl[ Q_{l\rightarrow i}^{5,h - 1}
      + Q_{l\rightarrow i}^{4, h}  \Bigr]
      \nonumber \\
      &\quad  \times \prod\limits_{m\in V \backslash (U_1\cup U_2)}
      \Bigl[ Q_{m\rightarrow i}^{0} +\sum\limits_{t=1}^{h-1} Q_{m\rightarrow i}^{3,t}
        +\sum\limits_{t=1}^{h-2} Q_{m\rightarrow i}^{5,t} \Bigr]
      \; ,
      \\
      G_{i}^{4, h}\bigl( x, y, V \bigr) & =
      \sum\limits_{U_1 \in V}^{|U_1| \leq x} \;
      \sum\limits_{U_2 \in V\backslash U_1}^{|U_1\cup U_2| \geq y} \;
      \prod\limits_{k\in U_1} \Bigl[ Q_{k\rightarrow i}^{2, h+1} +
        Q_{k\rightarrow i}^{5, h} + \sum\limits_{t \geq h+2}Q_{k\rightarrow i}^{1, t} \Bigr]
      \nonumber \\
      & \quad  \times 
      \prod\limits_{l\in U_2} \Bigl[ Q_{l\rightarrow i}^{5,h - 1} + Q_{l\rightarrow i}^{4, h}
        \Bigr]
      \nonumber \\
      & \quad \times \prod\limits_{m\in V \backslash (U_1\cup U_2)}
      \Bigl[ Q_{m\rightarrow i}^{0} +\sum\limits_{t=1}^{h-1} Q_{m\rightarrow i}^{3,t}
        +\sum\limits_{t=1}^{h-2} Q_{m\rightarrow i}^{5,t} \Bigr]
      \; ,
  \end{align}
\end{subequations}
where $\sum_{U_1\subset V}^{|U_1| \leq x}$ means summation over all the subsets $U_1$ of the vertex set $V$ under the constraint of the cardinality of $U_1$ not exceeding the integer value $x$, and $\sum_{U_2\subset V \backslash U_1}^{|U_1\cup U_2| \geq y}$ means summation over all the subsets $U_2$ of the vertex set $V\backslash U_1$ under the constraint of the cardinality of $U_1 \cup U_2$ being equal or larger than the integer value $y$. It is easy to check that $Q_{i\rightarrow j}^{3, H} = 0$.

The normalization constant $z_{i j}$ in the above coarse-grained BP equations is fixed by the requirement that 
\begin{eqnarray}
  & & (1+d_{j} H) Q_{i\rightarrow j}^{0}  + 2 Q_{i\rightarrow j}^{1, 1} 
  + \sum\limits_{h=1}^{H-1} (H-h) d_{j} Q_{i\rightarrow j}^{3,h}
  \nonumber \\
  & & \quad \quad +\sum\limits_{h=2}^{H}\Bigl[ \bigl( (h - 2) d_{j} +2 \bigr)
    Q_{i\rightarrow j}^{1,h} +d_{j} Q_{i\rightarrow j}^{2,h} \Bigr]
  \nonumber \\
  & & \quad \quad + \sum\limits_{h=1}^{H}\Bigl[ d_{j} Q_{i\rightarrow j}^{4,h}
    + \bigl( (H-h + 1) d_{j} -1 \bigr) Q_{i\rightarrow j}^{5,h} \Bigr]
  \  = \
  1 \; .
\end{eqnarray}

There are $(5 H-1)$ distinct coarse-grained cavity probabilities on each directed edge pointing from vertex $i$ to vertex $j$: $Q_{i\rightarrow j}^0$,  $Q_{i\rightarrow j}^{m, 1}$ (for $m=1, 3, 4, 5$), $Q_{i\rightarrow j}^{m, h}$ with $m=1,2,3,4,5$ and $h=2, \ldots, H$ (except for $m=3$ and $h=H$).  For a graph with $N$ vertices and $M$ undirected edges, we need to iterate these $(5 H-1)$ self-consistent equations on $2 M$ directed edges until they converge. The time complexity is then approximately of order $10 r H M$ with $r$ being the total number of BP iteration repeats. In our numerical simulation on the {\tt hCTGA} algorithm, we simply set $r$ to the minimum value $r = 1$.

A common and straightforward way to obtain a fixed-point solution of these coarse-grained BP equations on a single graph instance is by iteration on all the edges of the graph.  This works for low inverse temperature values. When the inverse temperature $\beta$ becomes large, it is often the case that the BP iteration fails to converge but instead becomes oscillatory. This non-convergence is often an indication of the emergence of spin glass phases~\cite{Mezard-2009}.  To partially suppress oscillatory behavior, we introduce a damping factor $\eta$ to the BP message-passing equations so that the old value of a cavity probability still contributes a weight $\eta$ to the new value of this cavity probability. For example, in the case of the cavity probability $Q_{i\rightarrow j}^{0}$, the modified version of Eq.~(\ref{eq:Qij0}) becomes
  \begin{equation}
    Q_{i\rightarrow j}^0 \
    \leftarrow \
    \eta Q_{i\rightarrow j}^0 + (1- \eta)
    \frac{e^{-\beta}}{z_{i\rightarrow j}}
    \prod\limits_{k\in \partial i\backslash j}\Bigl[ Q_{k\rightarrow i}^{0}
      + Q_{k\rightarrow i}^{1, 1} + \sum\limits_{h = 2}^H Q_{k\rightarrow i}^{1,h} \Bigr] \; .
  \end{equation}
  In the present paper we fix $\eta = 0.3$ in all the numerical simulations (the results are not sensitive to the precise value of $\eta$).

\subsection{
  Thermodynamic quantities
}
  
After reaching a fixed point of the coarse-grained BP equations, the free energy density $f$ is obtained by
\begin{equation}
  f\
  =
  \
  - \frac{1}{N \beta}\sum\limits_{i}^{N} \ln z_{i}  + \frac{1}{N \beta}
  \sum\limits_{(i,j)\in G} \ln z_{i j} \; ,
\end{equation}
where the two normalization constants $z_i$ and $z_{i j}$ are computed through
\begin{eqnarray}
  z_{i} & = & e^{-\beta } \prod\limits_{j\in \partial i} \Bigl[ Q_{j \rightarrow i}^{0}
    + Q_{j\rightarrow i}^{1, 1} + \sum\limits_{h=2}^{H} Q_{j\rightarrow i}^{1, h} \Bigr]
  + G_{i}^{1}\bigl( K-1,\; \partial i \bigr)
  \nonumber \\
  & & \quad  + \sum\limits_{j \in \partial i} Q_{j\rightarrow i}^{1,1} \;
  G_{i}^2 \bigl(K - 2,\; \partial i\backslash j \bigr)
  + \sum\limits_{h=2}^{H} G_{i}^{3, h} \bigl( K - 1 , \; K , \; \partial i \bigr)
  \nonumber \\
  & & \quad + \sum\limits_{h=2}^{H} \sum\limits_{j\in \partial i}
  Q_{j\rightarrow i}^{2, h}\; G_{i}^{4, h} \bigl( K - 2 ,\; K - 1 , \;
  \partial i\backslash j \bigr) \; ,
  \\
  z_{i j} & = & Q_{i\rightarrow j}^{0} Q_{j\rightarrow i}^{0}
  +Q_{i\rightarrow j}^{1, 1} Q_{j\rightarrow i}^{4,1}
  + Q_{i\rightarrow j}^{4, 1} Q_{j\rightarrow i}^{1,1}
  +\sum\limits_{h=1}^{H}\Bigl[ Q_{i\rightarrow j}^{1,h} Q_{j\rightarrow i}^{0}
    +Q_{i\rightarrow j}^{0} Q_{j\rightarrow i}^{1,h} \Bigr]
  \nonumber \\
  & & +\sum\limits_{h=2}^{H} \Bigl[ Q_{i\rightarrow j}^{4,h} Q_{j\rightarrow i}^{2,h}
    + Q_{i\rightarrow j}^{2,h} Q_{j\rightarrow i}^{4,h}\Bigr]
  +\sum\limits_{h=2}^{H}\sum\limits_{t=1}^{h-1}\Bigl[ Q_{i\rightarrow j}^{3,t}
    Q_{j\rightarrow i}^{1,h} +Q_{i\rightarrow j}^{1,h} Q_{j\rightarrow i}^{3,t}\Bigr]
  \nonumber \\
  & & + \sum\limits_{h=1}^{H} Q_{i\rightarrow j}^{5,h} Q_{j\rightarrow i}^{5,h}
  + \sum\limits_{h=3}^{H}\sum\limits_{t=1}^{h-2}
  \Bigl[ Q_{i\rightarrow j}^{1,h} Q_{j\rightarrow i}^{5,t}
    + Q_{i\rightarrow j}^{5,t} Q_{j\rightarrow i}^{1,h} \Bigr]
  \nonumber \\
  & &   
  +\sum\limits_{h=1}^{H-1}\Bigl[ Q_{i\rightarrow j}^{2,h+1} Q_{j\rightarrow i}^{5,h}
    +Q_{i\rightarrow j}^{5,h} Q_{j\rightarrow i}^{2,h+1} \Bigr]
  \; .
\end{eqnarray}
The marginal probability of vertex $i$ being a seed is then evaluated as
\begin{equation}
  q_{i}^{0}  = \frac{1}{z_{i}} e^{-\beta }
  \prod\limits_{j\in \partial i}\Bigl[ Q_{j\rightarrow i}^{0} + Q_{j\rightarrow i}^{1, 1} 
    + \sum\limits_{h=2}^{H} Q_{j\rightarrow i}^{1, h} \Bigr]
  \; ,
  \label{eq:qi0exp}
\end{equation}
and the mean energy density $\rho$ is
\begin{equation}
  \rho = \frac{1}{N}\sum_{i=1}^N  q_i^0 \;  .
\end{equation}
The entropy density at a fixed value of $\beta$ is determined as $s = \beta (\rho - f )$.

\subsection{
  Replica-symmetric mean field equations for regular random graph ensembles
}
\label{app:mfRR}

For the regular random graph ensemble with uniform vertex degree ($d_i = D$ for all the vertices $i$), we make the additional assumption that all the cavity probability messages are independent of the specific edges, namely $Q_{i\rightarrow j}^0 = Q^0$, $Q_{i\rightarrow j}^{1, h} = Q^{1, h}$, and so on. The normalization constant $z_{i \rightarrow j}$ will then be the same for all the edges $(i, j)$. The replica-symmetric mean field equations for the quantities $Q^0, Q^{1, h}, \ldots$ can be easily written down from the general coarse-grained BP equations. For layer $h = 0$ and $h = 1$, they are
\begin{subequations}
  \begin{align}
    Q^{0} & = \frac{1}{z_{i \rightarrow j}}
    e^{-\beta }
    \Bigl( Q^{0} + Q^{1, 1} + \sum\limits_{h=2}^{H} Q^{1, h} \Bigr)^{D-1} \; ,
    \\
    Q^{1,1} & = \frac{1}{z_{i \rightarrow j}} \biggl\{ \sum\limits_{n=0}^{K-1}
    \C{D-1}{n} \Bigl( \sum\limits_{h=2}^{H} Q^{1, h} \Bigr)^{n}
    \Bigl( Q^{0} +Q^{4,1} \Bigr)^{D-1-n} + (D-1)\; Q^{1,1}\; \times
    \nonumber    \\
    & \quad \quad 
    \sum\limits_{n=0}^{K-2}
    \C{D-2}{n}\Bigl( Q^{5,1} +Q^{2,2} +\sum\limits_{d=3}^{H} Q^{1,h}\Bigr)^{n}
    \Bigl( Q^{0} +Q^{4,1}\Bigr)^{D-2-n} \biggr\} \; ,
    \\
    Q^{3,1} & =  \frac{1}{z_{i \rightarrow j}}
    \sum\limits_{n=0}^{K-2} \C{D-1}{n} \Bigl( \sum\limits_{h=2}^{H} Q^{1,h}
    \Bigr)^{n} \Bigl( Q^{0} +Q^{4,1} \Bigr)^{D-1-n}
    \; ,
    \\
    Q^{4,1}  & =  \frac{1}{z_{i \rightarrow j}}
    \sum\limits_{n=0}^{K-2} \C{D-1}{n} \Bigl( Q^{5,1} +Q^{2,2}
    +\sum\limits_{h=3}^{H} Q^{1,h} \Bigr)^{n} \Bigl( Q^{0} +Q^{4,1} \Bigr)^{D-1-n}
    \; ,
    \\
    Q^{5,1}   & =  \frac{1}{z_{i \rightarrow j}}
    (D-1)\; Q^{1,1}\;
    \times
    \nonumber \\
    & \quad \quad \sum\limits_{n=0}^{K-3} \C{D-2}{n}
    \Bigl( Q^{5,1} +Q^{2,2} +\sum\limits_{h=3}^{H} Q^{1,h} \Bigr)^{n}
    \Bigl( Q^{0} +Q^{4,1} \Bigr)^{D-2-n}
    \; ,
  \end{align}
\end{subequations}
where $\C{D}{n} \equiv \frac{D !}{n!(D-n)!}$ is the binomial coefficient. For layer $h = 2, \ldots, H$, the self-consistent equations are
\begin{subequations}
  \begin{align}
    Q^{1, h}  & =  \frac{1}{z_{i \rightarrow j}} \biggl\{
    \sum\limits_{n=0}^{K-1}\; \sum\limits_{m=K-n}^{D-1-n}
    \C{D-1}{n} \C{D-1-n}{m} \Bigl(\sum\limits_{t=h+1}^{H} Q^{1,t} \Bigr)^{n}
    \Bigl( Q^{5,h-1} +Q^{4,h} \Bigr)^{m} \; \times
    \nonumber \\
    & \quad \quad \Bigl( Q^{0} +\sum\limits_{t=1}^{h-1} Q^{3,t}
    +\sum\limits_{t=1}^{h-2} Q^{5,t} \Bigr)^{D-1-n-m} + (D-1) Q^{2,h}\; \times
    \nonumber \\
    & \quad \quad \sum\limits_{n=0}^{K-2} \; \sum\limits_{m=K-1-n}^{D-2-n}
    \C{D-2}{n} \C{D-2-n}{m} \Bigl( Q^{2,h+1} +\sum\limits_{t=h+2}^{H} Q^{1,t}
    +Q^{5,h} \Bigr)^{n} \; \times
    \nonumber \\
    & \quad\quad \Bigl( Q^{5,h-1} +Q^{4,h} \Bigr)^{m}
    \Bigl( Q^{0} +\sum\limits_{t=1}^{h-1} Q^{3,t} +\sum\limits_{t=1}^{h-2}
    Q^{5,t} \Bigr)^{D-2-n-m} \biggr\} \; ,
    \\
    Q^{2,h}  & =  \frac{1}{z_{i \rightarrow j}} \biggl\{
    \sum\limits_{n=0}^{K-1}\; \sum\limits_{m=K-1-n}^{D-1-n} \C{D-1}{n} \C{D-1-n}{m}
    \Bigl(\sum\limits_{t=h+1}^{H} Q^{1,t}\Bigr)^{n} \Bigl( Q^{5,h-1} +Q^{4,h}\Bigr)^{m}
    \nonumber \\
    & \quad \quad \times \Bigl( Q^{0} +\sum\limits_{t=1}^{h-1} Q^{3,t}
    +\sum\limits_{t=1}^{h-2} Q^{5,t} \Bigr)^{D-1-n-m}
    +(D-1) Q^{2,h} \; \times
    \nonumber \\
    &\quad \quad \sum\limits_{n=0}^{K-2} \sum\limits_{m=K-2-n}^{D-2-n}
    \C{D-2}{n} \C{D-2-n}{m} \Bigl( Q^{2,h+1} +\sum\limits_{t=h+2}^{H} Q^{1,t}
    + Q^{5,h} \Bigr)^{n} \; \times
    \nonumber \\
    & \quad \quad \Bigl( Q^{5,h-1} +Q^{4,h} \Bigr)^{m}
    \Bigl( Q^{0} +\sum\limits_{t=1}^{h-1} Q^{3,t} +\sum\limits_{t=1}^{h-2} Q^{5,t}
    \Bigr)^{D-2-n-m} \biggr\} \; ,
    \\
    Q^{3,h}   & =  \frac{1}{z_{i \rightarrow j}}
    \sum\limits_{n=0}^{K-2}\; \sum\limits_{m=K-1-n}^{D-1-n} \C{D-1}{n} \C{D-1-n}{m}
    \Bigl(\sum\limits_{t=h+1}^{H} Q^{1,t}\Bigr)^{n} \Bigl( Q^{5,h-1} +Q^{4,h}\Bigr)^{m}
    \nonumber \\
    & \quad \quad \times \Bigl( Q^{0} +\sum\limits_{t=1}^{h-1}
    Q^{3,t} +\sum\limits_{t=1}^{h-2} Q^{5,t} \Bigr)^{D-1-n-m}
    \; ,
    \\
    Q^{4, h}   & =  \frac{1}{z_{i \rightarrow j}}
    \sum\limits_{n=0}^{K-2} \; \sum\limits_{m=K-1-n}^{D-1-n} \C{D-1}{n} \C{D-1-n}{m}
    \Bigl( Q^{2,h+1} +\sum\limits_{t=h+2}^{H} Q^{1,t} +Q^{5,h} \Bigr)^{n}
    \nonumber
    \\
    & \quad \quad \times \Bigl( Q^{5,h-1} +Q^{4,h} \Bigr)^{m}
    \Bigl( Q^{0} +\sum\limits_{t=1}^{h-1} Q^{3,t} +\sum\limits_{t=1}^{h-2} Q^{5,t}
    \Bigr)^{D-1-n-m}
    \; ,
    \\
    Q^{5,h}   & =  \frac{1}{z_{i \rightarrow j}}
    (D-1) Q^{2,h} \sum\limits_{n=0}^{K-3}\; \sum\limits_{m=K-2-n}^{D-2-n}
    \C{D-2}{n} \C{D-2-n}{m}  \; \times
    \nonumber \\
    & \quad \quad \Bigl( Q^{2,h+1} +\sum\limits_{t=h+2}^{H} Q^{1,t}
    +Q^{5,h} \Bigr)^{n} \Bigl( Q^{5,h-1} +Q^{4,h} \Bigr)^{m}
    \; \times
    \nonumber \\
    & \quad \quad
    \Bigl( Q^{0} +\sum\limits_{t=1}^{h-1} Q^{3,t} +\sum\limits_{t=1}^{h-2}
    Q^{5,t} \Bigr)^{D-2-n-m}
    \; .
  \end{align}
\end{subequations}
The normalization constant $z_{i \rightarrow j}$ is determined the condition
\begin{eqnarray}
  1 & = &  (1+D H) Q^{0} +2 Q^{1,1} + \sum\limits_{h=2}^{H} \Bigl[ \bigl( (h-2) D
    +2 \bigr) Q^{1,h} +D Q^{2,h} \Bigr] +
  \nonumber
  \\
  & &  \sum\limits_{h=1}^{H-1} (H-h) D Q^{3,h}
  + \sum\limits_{h=1}^{H} \Bigl[ D Q^{4,h} + \bigl( (H-h+1) D - 1 \bigr)
    Q^{5,h} \Bigr] \; .
\end{eqnarray}

At any given value of inverse temperature $\beta$, we can solve the coupled mean-field equations by the Newton-Raphson method or other numerical methods. At the fixed-point solution the thermodynamical quantities of interest can then be easily computed as follows. The average energy density $\rho$ (which is also the probability of a vertex belonging to the attack set) is
\begin{equation}
  \rho = \frac{1}{z_i} e^{-\beta }\Bigl( Q^{0} + Q^{1, 1} +
  \sum\limits_{h=2}^{H} Q^{1,h} \Bigr)^{D} \; ,
\end{equation}
where the normalization constant $z_i$ is
\begin{eqnarray}
  z_{i} &  = & e^{-\beta } \Bigl( Q^{0} + Q^{1, 1}
  + \sum\limits_{h=2}^{H} Q^{1,h}\Bigr)^{D}
  + \sum\limits_{n=0}^{K-1} \C{D}{n} \Bigl(\sum\limits_{h=2}^{H} Q^{1,h}\Bigr)^{n}
  \Bigl( Q^{0} +Q^{4,1} \Bigr)^{D-n}
  \nonumber \\
  & & + D Q^{1,1} \sum\limits_{n=0}^{K-2} \C{D-1}{n} \Bigl( Q^{5,1} + Q^{2,2}
  +\sum\limits_{h=3}^{H} Q^{1,h} \Bigr)^{n} \Bigl( Q^{0} +Q^{4,1} \Bigr)^{D-1-n}
  \nonumber \\
  & & +\sum\limits_{h=2}^{H} \sum\limits_{n=0}^{K-1} \; \sum\limits_{m=K-n}^{D-n}
  \C{D}{n} \C{D-n}{m} \Bigl( \sum\limits_{t=h+1}^{H} Q^{1,t} \Bigr)^{n}
  \Bigl( Q^{5,h-1} +Q^{4,h}\Bigr)^{m} \; \times
  \nonumber \\
  & & \quad  \Bigl( Q^{0} +\sum\limits_{t=1}^{h-1} Q^{3,t}
  +\sum\limits_{t=1}^{h-2} Q^{5,t} \Bigr)^{D-n-m} + D \sum\limits_{h=2}^{H} Q^{2,h}
  \; \times 
  \nonumber \\
  & & \quad \sum\limits_{n=0}^{K-2}\; \sum\limits_{m=K-1-n}^{D-1-n}
  \C{D-1}{n} \C{D-1-n}{m} \Bigl( Q^{2,h+1} +\sum\limits_{t=h+2}^{H} Q^{1,t} +Q^{5,h}
  \Bigr)^{n}
  \nonumber \\
  & & \quad \times \Bigl( Q^{5,h-1} +Q^{4,h} \Bigr)^{m} \Bigl( Q^{0}
  +\sum\limits_{t=1}^{h-1} Q^{3,t} +\sum\limits_{t=1}^{h-2} Q^{5,t} \Bigr)^{D-1-n-m}
  \; .
\end{eqnarray}
The free energy density $f$ is computed as
\begin{equation}
  f =  -\frac{1}{\beta } \ln z_{i} + \frac{D}{2\beta } \ln z_{i j} \; ,
\end{equation}
where the other normalization constant $z_{i j}$ is
\begin{eqnarray}
  z_{ij} & = &
  Q^{0} Q^{0} + 2 Q^0 Q^{1, 1} + 2 Q^{1,1} Q^{4,1} + 2Q^{0} \sum\limits_{h = 2}^{H} Q^{1,h}
  +2 \sum\limits_{h=2}^{H} Q^{2,h} Q^{4,h}
  \nonumber \\
  & & + 2 \sum\limits_{h=2}^{H} \sum\limits_{t=1}^{h-1} Q^{1,h} Q^{3,t}
  + 2 \sum\limits_{h=3}^{H}\sum\limits_{t=1}^{h-2} Q^{1,h} Q^{5, t}
  + 2 \sum\limits_{h=1}^{H-1} Q^{2, h+1} Q^{5,h}
  \nonumber \\
  & & 
  + \sum\limits_{h=1}^{H} Q^{5,h} Q^{5, h}
  \; .
\end{eqnarray}
The entropy density is $s = \beta ( \rho -f )$.

\section{Acknowledgments}

This work was supported by the National Natural Science Foundation of China (Grant Nos. 11975295, 12047503, and 12247104), the Chinese Academy of Sciences (Grant Nos. QYZDJ-SSW-SYS018, and XDPD15), and National Innovation Institute of Defense Technology Grant No. 22TQ0904ZT01025. Numerical simulations were carried out at the Tianwen cluster of ITP-CAS.

\section{Data availability statement}

The Python code used to generate the datasets of the provided examples is available from the first author, Jianwen Zhou (zhoujianwen@itp.ac.cn), on reasonable request.

\section{Conflict of interest statement}

We declared that we have no conflicts of interest in this work.


\end{document}